\documentclass[preprint]{elsarticle}
\usepackage[english]{babel}
\usepackage{graphicx}
\usepackage{dsfont}
\usepackage{bm}
\usepackage{amssymb}
\usepackage{amsthm}
\usepackage{amsmath}
\usepackage{amsfonts}
\usepackage{booktabs}
\usepackage{amsbsy}
\usepackage{fontenc}
\usepackage[utf8]{inputenc}
\usepackage{fancybox}
\usepackage[hypcap=false]{caption}
\usepackage{float}
\usepackage{subfigure} 
\usepackage{lmodern}
\usepackage[numbib,nottoc]{tocbibind}
\usepackage[pdfstartview={FitH},linkbordercolor={0 1 0}, colorlinks]{hyperref}
\usepackage{esint}
\usepackage{fancyhdr}
\usepackage{pdfpages}
\usepackage{trfsigns}
\usepackage{wasysym} 
\usepackage{siunitx} 
\usepackage{eqnarray} 
\usepackage{mathtools} 
\usepackage{relsize} 
\usepackage{scrhack} 
\usepackage{mleftright} 
\usepackage{tikz} 
\usetikzlibrary{cd}
\usepackage[many]{tcolorbox} 
\tcbuselibrary{theorems} 

\allowdisplaybreaks 

\renewcommand{\theequation}{\arabic{equation}}

\def\be{\begin{equation}}
\def\ee{\end{equation}}
\def\bs{\begin{subequations}}
\def\es{\end{subequations}}
\def\ba#1\ea{\begin{align}#1\end{align}}
\def\bes{\begin{equation*}}
\def\ees{\end{equation*}}
\def\bas#1\eas{\begin{align*}#1\end{align*}}

\theoremstyle{plain}
\newtheorem{theorem}{Theorem}[section]

\newtcbtheorem
  [use counter*=theorem,number within=section]
  {lemmata}
  {Lemma}
  {%
		breakable,
		enhanced,
    colback=gray!25,
    colframe=gray!0!black,
    fonttitle=\bfseries,
  }
  {lem}

\newtcbtheorem
  [use counter*=theorem,number within=section]
  {propositions}
  {Proposition}
  {%
		breakable,
		enhanced,
    colback=gray!25,
    colframe=gray!0!black,
    fonttitle=\bfseries,
  }
  {prop}
\newtcbtheorem
  [use counter*=theorem,number within=section]
  {theorems}
  {Theorem}
  {%
		breakable,
		enhanced,
    colback=gray!25,
    colframe=gray!0!black,
    fonttitle=\bfseries,
  }
  {thm}
\newtcbtheorem
  [use counter*=theorem,number within=section]
  {corollaries}
  {Corollary}
  {%
		breakable,
		enhanced,
    colback=gray!25,
    colframe=gray!0!black,
    fonttitle=\bfseries,
  }
  {cor}

\theoremstyle{remark}
\newtcbtheorem
  [use counter*=theorem,number within=section]
  {remarks}
  {Remark}
  {%
		breakable,
		enhanced,
    colback=gray!5,
    colframe=gray!50!black,
    fonttitle=\bfseries,
  }
  {rem}

\newtheorem{remark}[theorem]{Remarks}
\newtheorem*{remarkohne}{Remarks}

\theoremstyle{definition}

\newtcbtheorem
  [use counter*=theorem,number within=section]
  {definitions}
  {Definition}
  {%
		breakable,
		enhanced,
    colback=gray!25,
    colframe=gray!0!black,
    fonttitle=\bfseries,
  }
  {def}
\newtcbtheorem
  [use counter*=theorem,number within=section]
  {examples}
  {Example}
  {
		breakable,
		enhanced,
    colback=gray!5,
    colframe=gray!50!black,
    fonttitle=\bfseries,
  }
  {ex}
\newtcbtheorem
  [use counter*=theorem,number within=section]
  {situations}
  {Situation}
  {
		breakable,
		enhanced,
    colback=gray!5,
    colframe=gray!50!black,
    fonttitle=\bfseries,
  }
  {sit}
\newtcbtheorem
  [use counter*=theorem,number within=section]
  {fieldredefinitions}
  {Theorem}
  {
		breakable,
		enhanced,
    colback=gray!25,
    colframe=gray!0!black,
    fonttitle=\bfseries,
  }
  {fieldredef}

\DeclareFontFamily{U}{mathx}{\hyphenchar\font45}
\DeclareFontShape{U}{mathx}{m}{n}{
      <5> <6> <7> <8> <9> <10>
      <10.95> <12> <14.4> <17.28> <20.74> <24.88>
      mathx10
      }{}
\DeclareSymbolFont{mathx}{U}{mathx}{m}{n}
\DeclareFontSubstitution{U}{mathx}{m}{n}
\DeclareMathAccent{\widecheck}{0}{mathx}{"71}
\DeclareMathAccent{\wideparen}{0}{mathx}{"75}

\begin{document}
\pagenumbering{Roman}

\begin{frontmatter}

\title{Curved Yang-Mills-Higgs gauge theories in the case of massless gauge bosons\tnoteref{mytitlenote}}
\tnotetext[mytitlenote]{Abbreviations used in this paper: \textbf{(C)YMH GT} for (curved) Yang-Mills-Higgs gauge theory, and \textbf{LAB} for Lie algebra bundle.}

\author[mymainaddress,mysecondaryaddress]{Simon-Raphael Fischer\corref{myfootnote}}
\cortext[myfootnote]{Email: \href{mailto:Simon-Raphael.Fischer@unige.ch}{Simon-Raphael.Fischer@unige.ch}, \href{mailto:sfischer@math.univ-lyon1.fr}{sfischer@math.univ-lyon1.fr}.}



\address[mymainaddress]{Villa Battelle, Université de Genève, Route de Drize 7, 1227 Carouge, Switzerland}
\address[mysecondaryaddress]{Institut Camille Jordan, Université Claude Bernard Lyon 1, 43 boulevard du 11 novembre 1918, 69622 Villeurbanne cedex, France}

\begin{abstract}
Alexei Kotov and Thomas Strobl have introduced a covariantized formulation of Yang-Mills-Higgs gauge theories whose main motivation was to replace the Lie algebra with Lie algebroids. This allows the introduction of a possibly non-flat connection $\nabla$ on this bundle, after also introducing an additional 2-form $\zeta$ in the field strength. We will study this theory in the simplified situation of Lie algebra bundles, \textit{i.e.}~only massless gauge bosons, and we will provide a physical motivation of $\zeta$. Moreover, we classify $\nabla$ using the gauge invariance, resulting into that $\nabla$ needs to be a Lie derivation law covering a pairing $\Xi$, as introduced by Mackenzie. There is also a field redefinition, keeping the physics invariant, but possibly changing $\zeta$ and the curvature of $\nabla$. We are going to study whether this can lead to a classical theory, and we will realize that this has a strong correspondence to Mackenzie's study about extending Lie algebroids with Lie algebra bundles. We show that Mackenzie's obstruction class is also an obstruction for having non-flat connections which are not related to a flat connection using the field redefinitions. This class is related to $\mathrm{d}^\nabla \zeta$, a tensor which also measures the failure of the Bianchi identity of the field strength and which is invariant under the field redefinition. This tensor will also provide hints about whether $\zeta$ can vanish after a field redefinition.
\end{abstract}

\begin{keyword}
\texttt{Mathematical Gauge Theory}\sep Differential Geometry\sep High Energy Physics - Theory\sep Mathematical Physics
\MSC[2020] 53D17\sep 81T13\sep 17B99 \\
\textit{Subject Classification:} classical field theory\sep Lie groups and Lie (super)algebras\sep symplectic geometry \\
\textit{DOI:} \href{https://doi.org/10.1016/j.geomphys.2021.104104}{10.1016/j.geomphys.2021.104104}\\
\textit{License:} © 2021. This manuscript version is made available under the CC-BY-NC-ND 4.0 license \url{http://creativecommons.org/licenses/by-nc-nd/4.0}
\end{keyword}

\end{frontmatter}



\newpage


\tableofcontents

\setlength{\parindent}{12 pt}


\pagenumbering{arabic}

\section{Introduction}

The research concerns curved Yang-Mills-Higgs gauge theories (short: \textbf{CYMH GT}), introduced by Alexei Kotov and Thomas Strobl, where essentially the structural Lie algebra together with its action on the manifold $N$ of values of the Higgs field is replaced by a general Lie algebroid $E \to N$. We introduce Lie algebroids later, but one possible difference is that we now have structure functions instead of constants as usually in gauge theory and particle physics. Moreover, this Lie algebroid is equipped with a connection, a metric on the base, a fibre metric on $E$, and, last but not least, a 2-form $\zeta$ on $N$ with values in $E$ which contributes to the field strength. Gauge invariance of the Yang-Mills type functional leads to several \textbf{compatibility conditions} to be satisfied between these structures. When the connection $\nabla$ on $E$ is  flat, the compatibilities imply that the Lie algebroid is locally a so-called action Lie algebroid, and one gets back to the standard Yang-Mills-Higgs gauge theory when additionally $\zeta \equiv 0$. Thus, the theory represents a covariantized version of gauge theory equipped with an additional 2-form $\zeta$. When $\nabla$ is flat we say in general that we have a \textbf{pre-classical} gauge theory, and when additionally $\zeta \equiv 0$ we have a \textbf{classical} gauge theory.

The 2-form $\zeta$ is needed to allow non-flat connections, because otherwise only flat connections could satisfy the compatibility conditions. But $\zeta$ may not just be an auxiliary map, there is also a field redefinition for the classical formulation of gauge theory which keeps the Lagrangian invariant but it is adding this 2-form to the existing theory. It is then natural to study a gauge theory where $\zeta$ is non-zero and cannot be transformed to zero by the mentioned field redefinition. This field redefinition leads also to transformations of some other data, without violating the compatibility conditions. Most noteworthy, it also permits to change the curvature of $\nabla$ and not only to change or to create $\zeta$. The main motivation of this paper is to answer the question when such a field redefinition leads to a pre-classical or even classical theory, but in the simplified setting of using Lie algebra bundles and not general Lie algebroids.

After some basic definitions in \S \ref{BasicDefinitions}, we state in \S \ref{CYMHGTWITHLABS} what CYMH GT explicitly means in the simplified situation using just Lie algebra bundles $K \to N$, which means that we only consider massless gauge bosons for simplicity. Afterwards we discuss the field redefinitions in that situation, and provide the mentioned physical motivation for $\zeta$. Hereafter, we start with a discussion about whether there is a field redefinition making $\nabla$ flat:

In \S \ref{ConnectionIsALieDerivation} 
and \S \ref{MackenzieZeugsUndExistenzvonPreclassical} we will see that the question about whether we have a field redefinition transforming the gauge theory into a pre-classical one, has a strong relation to Mackenzie's study about extending Lie algebroids with Lie algebra bundles, and the obstruction of extensions of Lie algebroids by Lie algebra bundles have a relation to the obstruction of CYMH GTs; this obstruction will be related to $\mathrm{d}^\nabla \zeta$, an invariant of the field redefinition.
This relation allows to answer a question, coming from a physical context, with the results about extensions of Lie algebroids.
%
%
%

In \S \ref{NonclassicalStuff} we finally start discussing whether there is a field redefinition making $\zeta$ zero.
This is also again related to $\mathrm{d}^\nabla \zeta$: With the condition $\mathrm{d}^\nabla \zeta \neq 0$ we are able to show that there is then no field redefinition leading to a vanishing $\zeta$. Additionally, we provide a canonical construction of a gauge theory with $\mathrm{d}^\nabla \zeta \neq 0$, starting with a standard/classical Yang-Mills gauge theory.

Finally, in \S \ref{BianchiStuff}, we turn to the discussion about a possible physical meaning of $\mathrm{d}^\nabla \zeta \neq 0$. We are going to see that it measures the failure of the Bianchi identity of the new field strength, \textit{i.e.}~$\mathrm{d}^\nabla \zeta = 0$ if and only if the Bianchi identity is satisfied.
\section{Basic definitions}\label{BasicDefinitions}

In the following, we denote with $V^*$ the dual of a vector bundle $V \to N$ over a smooth manifold $N$, and $X^*V$ denotes the pull-back of $V$ by $X: M \to N$, a smooth map from a smooth manifold $M$ to $N$. We have a similar notation for the pull-back of sections, especially we will have sections $F$ as an element of $\Gamma\left( \left(\bigotimes_{m=1}^{l} E_m^*\right) \otimes E_{l+1} \right)$, where $E_1, \dots E_{l+1} \to N$ ($l \in \mathbb{N}$) are real vector bundles of finite rank over a smooth manifold $N$, and $\Gamma(\cdot)$ denotes the space of smooth sections. Then we view the pull-back $X^*F$ as an element of $\Gamma\left( \mleft(\bigotimes_{m=1}^{l} \mleft(X^*E_m\mright)^*\mright) \otimes X^*E_{l+1} \right)$, and it is essentially given by
\bas
	(X^*F)(X^*\nu_1, \dotsc , X^*\nu_l)
	&=
	X^*\mleft( F\mleft( \nu_1, \dotsc, \nu_l \mright) \mright)
\eas
for all $\nu_1 \in \Gamma(E_1), \dotsc, \nu_l \in \Gamma(E_l)$. In general we also make use of that sections of $X^*E$ can be viewed as sections of $E$ along $X$, where $E \stackrel{\pi}{\to} N$ is any vector bundle over $N$. Let $\mu \in \Gamma(X^*E)$, then it has the form $\mu_p = (p, u_p)$ for all $p \in M$, where $u_p \in E_{X(p)}$, the fibre of $E$ at $X(p)$; and a section $\nu$ of $E$ along $X$ is a smooth map $M \to E$ such that $\pi \circ \nu = X$. Then on one hand $\mathrm{pr}_2 \circ \mu$ is a section along $X$, where $\mathrm{pr}_2$ is the projection onto the second component, and on the other hand $M \ni p \mapsto (p, \nu_p)$ defines an element of $\Gamma(X^*E)$. With that one can show that there is a 1:1 correspondence of $\Gamma(X^*E)$ with sections along $X$. We do not necessarily mention it when we make use of that identification, it should be clear by the context.

Additionally, with $\Omega^k(N; E)$ ($k \in \mathbb{N}_0$) we denote $k$-forms on $N$ with values in a vector bundle $E \to N$, and recall the following wedge product\footnote{As also defined in \cite[\S 5, third part of Exercise 5.15.12; page 316]{hamilton}.} of forms with values in a vector bundle $E$ and values in its space of endomorphisms $\mathrm{End}(E)$,
\bas
\wedge: \Omega^k(N; \mathrm{End}(E)) \times \Omega^l(N; E)
&\mapsto
\Omega^{k+l}(N; E) \\
(T, \omega) &\mapsto T \wedge \omega
\eas
for all $k, l \in \mathbb{N}_0$, given by
\ba\label{DefVonWedgedemitEnd}
\mleft( T \wedge \omega \mright) \mleft( Y_1, \dotsc, Y_{k+l} \mright)
&\coloneqq
\frac{1}{k! l!} \sum_{\sigma \in S_{k+l}} \mathrm{sgn}(\sigma) ~
	T \mleft( Y_{\sigma(1)}, \dotsc, Y_{\sigma(k)} \mright)
		\mleft( \omega\mleft( Y_{\sigma(k+1)}, \dotsc, Y_{\sigma(k+l)} \mright) \mright),
\ea
where $S_{k+l}$ is the group of permutations $\{1, \dotsc, k+l\}$. This is then locally given by, with respect to a frame $\mleft( e_a \mright)_a$ of $E$,
\bas
T \wedge \omega &= T(e_a) \wedge w^a,
\eas
where $T$ acts as an endomorphism on $e_a$, \textit{i.e.}~$T(e_a) \in \Omega^k(N; E)$, and $\omega = \omega^a \otimes e_a$. Also recall that there is the canonical extension of $\nabla$ on $\mathrm{End}(E)$ by forcing the Leibniz rule. We still denote this connection by $\nabla$, too.

We also need the following definitions.

\begin{definitions}{Graded extension of products}{GradingOfProducts}
Let $l \in \mathbb{N}$, $E_1, \dots E_{l+1} \to N$ be real vector bundles of finite rank over a smooth manifold $N$, and $F \in \Gamma\left( \left(\bigotimes_{m=1}^{l} E_m^*\right) \otimes E_{l+1} \right)$. Then we define the \textbf{graded extension of $F$} as
	\bas
\Omega^{k_1}(N; E_1) \times \dots \times \Omega^{k_l}(N; E_l)
&\to \Omega^{k}(N; E_{l+1}), \\
(A_1, \dots, A_l)
&\mapsto
F\mleft(A_1\stackrel{\wedge}{,} \dotsc \stackrel{\wedge}{,} A_l\mright),
\eas
where $k := k_1+\dots k_l$ and $k_i \in \mathbb{N}_0$ for all $i\in \{1, \dots, l\}$. $F\mleft(A_1\stackrel{\wedge}{,} \dotsc \stackrel{\wedge}{,} A_l\mright)$ is defined as an element of $\Omega^{k}(N; E_{l+1})$ by
\bas
&F\mleft(A_1\stackrel{\wedge}{,} \dotsc \stackrel{\wedge}{,} A_l\mright)\mleft(Y_1, \dots, Y_{k}\mright)
\coloneqq \\
&\frac{1}{k_1! \cdot \dots \cdot k_l!} \sum_{\sigma \in S_{k}} \mathrm{sgn}(\sigma) ~ F\left( A_1\left( Y_{\sigma(1)}, \dots, Y_{\sigma(k_1)} \right), \dots, A_l\left( Y_{\sigma(k-k_l+1)}, \dots, Y_{\sigma(k)} \right) \right)
\eas
for all $Y_1, \dots, Y_{k} \in \mathfrak{X}(N)$, where $S_{k}$ is the group of permutations of $\{1, \dots, k\}$ and $\mathrm{sgn}(\sigma)$ the signature of a given permutation $\sigma$. 

$\stackrel{\wedge}{,}$ may be written just as a comma when a zero-form is involved.

Locally, with respect to given frames $\mleft( e^{(i)}_{a_i} \mright)_{a_i}$ of $E_i$, this definition has the form
\bas
F\mleft(A_1\stackrel{\wedge}{,} \dotsc \stackrel{\wedge}{,} A_l\mright)
&=
F\mleft(e^{(1)}_{a_1}, \dotsc, e^{(l)}_{a_l}\mright) \otimes A_1^{a_1} \wedge \dotsc \wedge A_l^{a_l}
\eas
for all $A_i = A_i^{a_i} \otimes e^{(i)}_{a_i}$, where $A_i^{a_i}$ are $k_i$-forms. 
\end{definitions}

\begin{remark}
\leavevmode\newline
$\bullet$ Assume $F \in \Gamma\left( \mleft(\bigwedge_{m=1}^{l} \mathrm{T}^*N \mright) \otimes E \right) \cong \Omega^l(N; E)$ for some vector bundle $E$, \textit{i.e.}~$F$ is an $l$-form on $N$ with values in $E$. Let $M$ be another smooth manifold, then the pull-back $X^*F$ by a smooth function $X \in C^\infty(M;N)$ can be viewed as an element of $\Gamma\left( \bigwedge_{m=1}^{l} \mleft(X^*\mathrm{T}N\mright)^* \otimes X^*E \right)$.

Do not confuse this pull-back with the pull-back of forms, here denoted by $X^!F$, which is an element of $\Gamma\left( \mleft(\bigwedge_{m=1}^{l} \mathrm{T}^*M \mright) \otimes X^*E \right) \cong \Omega^l(M; X^*E)$ defined by
\ba
\mleft.\mleft(X^!F\mright)(Y_1, \dots, Y_l)\mright|_p
&\coloneqq
F_{X(p)}\mleft(\mathrm{D}_pX\mleft(\mleft.Y_1\mright|_p\mright), \dots, \mathrm{D}_pX\mleft(\mleft.Y_l\mright|_p\mright)\mright)
\ea
for all $p \in M$ and $Y_1, \dots, Y_l \in \mathfrak{X}(M)$, where $\mathrm{D}X$ is the total differential of $X$ (also called tangent map). In the following we view $\mathrm{D}X$ as an element of $\Omega^1(M; X^*\mathrm{T}N)$ by $\mathfrak{X}(M) \ni Y \mapsto \mathrm{D}X(Y)$, where $\mathrm{D}X(Y) \in \Gamma(X^*\mathrm{T}N), M \ni p \mapsto \mathrm{D}_pX(Y_p)$. Then
\ba\label{EqPullBackFormelFuerVerschiedeneDefinitionen}
X^!F 
&=
\frac{1}{l!}~
\mleft(X^*F\mright) ( \underbrace{\mathrm{D}X \stackrel{\wedge}{,} \dotsc \stackrel{\wedge}{,} \mathrm{D}X}_{l \text{ times}} )
\ea
by using the anti-symmetry of $F$ and Def.~\ref{def:GradingOfProducts}, \textit{i.e.}
\bas
&\mleft.\frac{1}{l!}~
\Big(\mleft(X^*F\mright) ( \mathrm{D}X \stackrel{\wedge}{,} \dotsc \stackrel{\wedge}{,} \mathrm{D}X ) \Big) (Y_1, \dots, Y_l)\mright|_p \\
&\hspace{1cm}
=
\frac{1}{l!}~
\sum_{\sigma \in S_{l}} \mathrm{sgn}(\sigma) ~ \underbrace{(X^*F)\mleft(\mathrm{D}X\mleft(Y_{\sigma(1)}\mright), \dots, \mathrm{D}X\mleft(Y_{\sigma(l)}\mright)\mright)}_{\mathclap{= \mathrm{sgn}(\sigma) ~ (X^*F)\mleft(\mathrm{D}X\mleft(Y_{1}\mright), \dots, \mathrm{D}X\mleft(Y_{l}\mright)\mright)}}\Big|_p \\
&\hspace{1cm}
=
\frac{1}{l!}~ \underbrace{\mleft( \sum_{\sigma \in S_{l}} 1 \mright)}_{= l!} ~
F_{X(p)}\mleft(\mathrm{D}_pX\mleft(\mleft.Y_{1}\mright|_p\mright), \dots, \mathrm{D}_pX\mleft(\mleft.Y_{l}\mright|_{p}\mright)\mright) \\
&\hspace{1cm}
= \mleft.\mleft(X^!F\mright)(Y_1, \dots, Y_l)\mright|_p
\eas
for all $p \in M$ and $Y_1, \dots, Y_l \in \mathfrak{X}(M)$.

$\bullet$ This generalizes the expression for $\mleft[ A \stackrel{\wedge}{,} A \mright]_{\mathfrak{g}}$ often given in gauge theory for $A \in \Omega^1(N; \mathfrak{g})$ as \textit{e.g.}~defined in \cite[\S 5, Definition 5.5.3; page 275]{hamilton}, where $\mleft[ \cdot, \cdot \mright]_{\mathfrak{g}}$ is the Lie bracket of a Lie algebra $\mathfrak{g}$. To observe this, take a trivial Lie algebra bundle $N \times \mathfrak{g} \to N$ with a field of Lie brackets $\mleft[ \cdot, \cdot\mright]_K \in \Gamma\mleft( \bigwedge^2 K^* \otimes K \mright)$, which restricts to $\mleft[ \cdot, \cdot \mright]_{\mathfrak{g}}$ on each fibre. Take a a constant frame $\mleft( e_a \mright)_a$ of $K$ and view $A$ as an element of $\Omega^1(N; K)$, then
\bas
\mleft.\mleft[ A \stackrel{\wedge}{,} A\mright]_K(Y, Z)\mright|_p
&=
\mleft[ A_p(Y_p), A_p(Z_p) \mright]_{\mathfrak{g}}
 - \mleft[ A_p(Z_p), A_p(Y_p) \mright]_{\mathfrak{g}}
=
2~\mleft[ A_p(Y_p), A_p(Z_p) \mright]_{\mathfrak{g}}
\eas
for all $Y, Z \in \mathfrak{X}(N)$ and $p \in N$, and locally
\ba\label{eqKlammervonFormenInKoord}
\mleft[ A \stackrel{\wedge}{,} A\mright]_K
&=
\mleft[ e_a, e_b \mright]_{\mathfrak{g}} \otimes A^a \wedge A^b
\ea
by writing $A = A^a \otimes e_a$, where we view sections of $K$ as smooth maps $N \to \mathfrak{g}$.
\end{remark}

We also need to know what a Lie algebroid is, a generalization of both, tangent bundles and Lie algebras; this concept will just be defined, refer to the references for thorough discussions of these definitions, especially \cite{mackenzieGeneralTheory} and \cite[\S VII; page 113ff.]{DaSilva}. For the final statements and the context it is not necessarily important to know what a general Lie algebroid is.

\begin{definitions}{Lie algebroid, \newline \cite[\S 3.3, first part of Definition 3.3.1; page 100]{mackenzieGeneralTheory}}{test}
Let $E \to N$ be a real vector bundle of finite rank. Then $E$ is a smooth Lie algebroid if there is a bundle map $\rho: E \to \mathrm{T}N$, called the \textbf{anchor}, and a Lie algebra structure on $\Gamma(E)$ with Lie bracket $\mleft[ \cdot, \cdot \mright]_E$ satisfying
\ba
  \mleft[\mu, f \nu\mright]_E = f \mleft[\mu, \nu\mright]_E + \rho(\mu)(f) ~ \nu
\label{eq:E-Leibniz}
\ea
for all $f \in C^\infty(N)$ and $\mu, \nu \in \Gamma(E)$, where $\rho(\mu)(f)$ is the action of the vector field $\rho(\mu)$ on the function $f$ by derivation. We will sometimes denote a Lie algebroid by $\mleft( E, \rho, \mleft[ \cdot, \cdot \mright]_E \mright)$.
\end{definitions}

Tangent bundles are a canonical example of Lie algebroids, their anchor is the identity with which we also equip them; another canonical example with zero anchor are the Lie algebra bundles:

\begin{definitions}{Lie algebra bundle (LAB), \cite[Definition 3.3.8; page 104]{mackenzieGeneralTheory}}{LAB}
Let $\mathfrak{g}$ be a Lie algebra. A \textbf{Lie algebra bundle}, or \textbf{LAB}, is a vector bundle $K \to N$ equipped with a field of Lie algebra brackets $\mleft[ \cdot, \cdot \mright]_{K}: \Gamma(K) \times \Gamma(K) \to \Gamma(K)$, \textit{i.e.}~$\mleft[ \cdot, \cdot \mright]_{K} \in \Gamma\mleft(\bigwedge^2 K^* \otimes K \mright)$ such that it restricts to the Lie algebra bracket of $\mathfrak{g}$ on each fibre, and such that $K$ admits an atlas $\{ \psi_i: K|_{U_i} \to U_i \times \mathfrak{g} \}$ subordinate to some open covering $\mleft( U_i \mright)_i$ of $N$ in which each induced $\psi_{i, p}: K_p \to \mathfrak{g}$ is a Lie algebra isomorphism, where $p \in U_i$, $K_p$ the fiber at $p$, $\psi_{i, p} \coloneqq \mathrm{pr}_2 \circ \mleft.\psi_i\mright|_{K_p}$ and $\mathrm{pr}_2$ is the projection onto the second factor.
\end{definitions}

We need to know a special kind of morphism of Lie algebroids over the same base.

\begin{definitions}{Base-preserving morphism of Lie algebroids, \newline \cite[\S 3.3, second part of Definition 3.3.1; page 100]{mackenzieGeneralTheory}}{BasePreservingMorphismOfLieAlgebroids}
Let $\mleft( E_1, \rho_{E_1}, \mleft[ \cdot, \cdot \mright]_{E_1} \mright)$ and $\mleft( E_2, \rho_{E_2}, \mleft[ \cdot, \cdot \mright]_{E_2} \mright)$ be two Lie algebroids over the same base ma-nifold $N$. Then a \textbf{morphism of Lie algebroids $\phi: E_1 \to E_2$ over $N$}, or a \textbf{base-preserving morphism of Lie algebroids}, is a vector bundle morphism with
\bas
\rho_{E_2} \circ \phi &= \rho_{E_1}, \\
\phi\mleft( \mleft[ \mu, \nu \mright]_{E_1} \mright) &= \mleft[ \phi(\mu), \phi(\nu) \mright]_{E_2}
\eas
for all $\mu, \nu \in \Gamma(E_1)$.

When $\phi$ is additionally an isomorphism of vector bundles then we call it an \textbf{isomorphism of Lie algebroids over $N$}, or a \textbf{base-preserving isomorphism of Lie algebroids}.
\end{definitions}

As it can be seen by this definition, when we speak about bundle morphisms, then we always mean base-preserving ones, even when we do not explicitly mention this.

Related to some of these and following definitions we need some identities for the following. These are generalizations of many identities known in standard gauge theory and/or related to pull-backs. The identities and their proofs are very natural and straightforward, which is the reason why you can find them in \ref{CalculusIdentitiesNeeded}.

%
\section{(Curved) Yang-Mills-Higgs gauge theories on Lie algebra bundles} \label{CYMHGTWITHLABS}

Let us now start to look at covariantized gauge theories in the situation of Lie algebra bundles. What does it mean to have a (curved) Yang-Mills-Higgs gauge theory, as \textit{e.g.}~introduced in \cite[and the references therein]{CurvedYMH}? With gauge theory we just mean the infinitesimal gauge transformations, and the definitions are motivated by classical gauge theories with trivial principal bundle. Let us summarise the theory of Alexei Kotov and Thomas Strobl in the situation of LABs, whose physical context is then given by a Yang-Mills-Higgs gauge theory with just massless gauge bosons. The latter is the reason why we call the Higgs field $X$ just as an \textbf{additional free physical field} (here always given by a Lagrangian similar to the Higgs field, but without minimal coupling to the gauge bosons). In Remark~\ref{rem:MotivationForTheNewFieldStrength}, we give another motivation about deriving this theory, different to the original one.

\begin{situations}{CYMH GT for Lie algebra bundles, \newline \cite[but here a simplified and coordinate-free version in the setting of Lie algebra bundles]{CurvedYMH}}{CYMHGTForLABsToDoList}
Let $\mathfrak{g}$ be a real finite-dimensional Lie algebra with Lie bracket $\mleft[ \cdot, \cdot \mright]_{\mathfrak{g}}$. With
\begin{center}
	\begin{tikzcd}
		\mathfrak{g} \arrow{r}	& \mleft(K, \mleft[ \cdot, \cdot \mright]_K\mright) \arrow{d} \\
			& N
	\end{tikzcd}
\end{center}
we denote an LAB over a smooth manifold $N$ with Lie algebra structure inherited by $\mathfrak{g}$, with its field $\mleft[ \cdot, \cdot \mright]_K \in \Gamma\mleft(\bigwedge^2 K^* \otimes K \mright)$ of Lie brackets which restricts to the Lie bracket $\mleft[ \cdot, \cdot \mright]_{\mathfrak{g}}$ on each fibre.
%

Let $(M, \eta)$ be a spacetime $M$ with its spacetime metric $\eta$, and $X: M \to N$ a smooth map. For defining terms like the field strength we need to have a Lie algebra bundle structure over $M$ itself and therefore we look at the vector bundle pull-back $X^*K$. $X^*K$ has also the structure of an LAB with a field of Lie brackets denoted by $\mleft[ \cdot, \cdot \mright]_{X^*K} \in \Gamma\mleft(\bigwedge^2 X^*\mleft(K^*\mright) \otimes X^*K \mright)$, which restricts to $\mleft[ \cdot, \cdot \mright]_{\mathfrak{g}}$ on each fibre, too. This bracket is given by
\bas
\mleft[ \cdot, \cdot \mright]_{X^*K}
&=
X^*\mleft(\mleft[ \cdot, \cdot \mright]_{K}\mright),
\eas
where we view $\mleft[ \cdot, \cdot \mright]_{K}$ as a section when making the pull-back.
Hence, we arrive at:
\begin{center}
	\begin{tikzcd}
		  \mleft(X^*K, \mleft[ \cdot, \cdot \mright]_{X^*K}\mright) \arrow{d} & \mleft(K, \mleft[ \cdot, \cdot \mright]_K\mright) \arrow{d} \\
			(M, \eta) \arrow{r}{X} & N
	\end{tikzcd}
\end{center}
Let us also fix a vector bundle connection $\nabla$ on $K$, and therefore also a connection on $X^*K$ by using the definition of the pull-back connection, $X^*\nabla$. We also have $\zeta \in \Omega^2(N; K)$ such that
\ba\label{CondSGleichNullLAB}
\nabla_Y\mleft( \mleft[ \mu, \nu \mright]_K \mright)
&=
\mleft[ \nabla_Y \mu, \nu \mright]_K
	+ \mleft[ \mu, \nabla_Y \nu \mright]_K, \\
R_\nabla(Y, Z) \mu
&=
\mleft[ \zeta(Y, Z), \mu \mright]_K \label{CondKruemmungmitBLAB}
\ea
for all $Y, Z \in \mathfrak{X}(N)$ and $\mu, \nu \in \Gamma(K)$, where $R_\nabla$ is the curvature of $\nabla$. These are the \textbf{compatibility conditions}.

The \textbf{field of gauge bosons} will be represented by
\bas
A &\in \Omega^1(M; X^*K)
\eas
\textit{i.e.}~a one-form on $M$ with values in $X^*K$. The \textbf{field strength} $G$ is then defined as an element of $\Omega^2(M; X^*K)$ by
\ba
G_{A,X}
\coloneqq
G
&\coloneqq
\mathrm{d}^{X^*\nabla}A
	+ \frac{1}{2} \mleft[ A \stackrel{\wedge}{,} A \mright]_{X^*K}
	+ \frac{1}{2} \mleft( X^*\zeta \mright)\mleft( \mathrm{D}X \stackrel{\wedge}{,} \mathrm{D}X \mright) \nonumber \\
&=
\mathrm{d}^{X^*\nabla}A
	+ \frac{1}{2} \mleft[ A \stackrel{\wedge}{,} A \mright]_{X^*K}
	+ X^!\zeta, \label{defNewFieldStrengthG}
\ea
where $\mathrm{d}^{X^*\nabla}$ is the exterior covariant derivative with respect to $X^*\nabla$.

Due to \eqref{CondSGleichNullLAB} we have the Lie algebra of \textbf{infinitesimal gauge transformations}, these are parametrised by sections $\varepsilon \in \Gamma(X^*K)$ and given by
\ba\label{EqInfGaugeTrafoLABs}
\delta_\varepsilon A
&\coloneqq
	\mleft[ \varepsilon, A \mright]_{X^*K}
		- \mathrm{d}^{X^*\nabla} \varepsilon, \\
\delta_\varepsilon X &\coloneqq 0.
\ea
As usual, the infinitesimal gauge transformation $\delta_\varepsilon G$ of $G$ is defined by
\ba
\delta_\varepsilon G
\coloneqq
\mleft.\frac{\mathrm{d}}{\mathrm{d}t}\mright|_{t=0} G_{A + t \cdot \delta_\varepsilon A, X}
\ea
for $t \in \mathbb{R}$. Because of the compatibility conditions~\eqref{CondSGleichNullLAB} and~\eqref{CondKruemmungmitBLAB} we can derive that $\delta_\varepsilon G$ has the following form
\ba
\delta_\varepsilon G
&=
\mleft[ \varepsilon, G \mright]_{X^*K}.
\ea

The (uncoupled) Yang-Mills-Higgs Lagrangian $\mathcal{L}_{\mathrm{YMH}}$ of $A$ and $X$ is then defined as a top-degree-form $\mathcal{L}_{\mathrm{YMH}}[A, X] \in \Omega^{\mathrm{dim}(M)}(M)$ given by 
\ba\label{defLagrangianForLABs}
\mathcal{L}_{\mathrm{YMH}}[A, X]
&\coloneqq
- \frac{1}{2} \mleft( X^*\kappa \mright)(G \stackrel{\wedge}{,} *G)
	+ \mleft( X^*g \mright)(\mathrm{D}X \stackrel{\wedge}{,} *\mathrm{D}X)
	+ *(V \circ X),
\ea
where $*$ is the Hodge star operator w.r.t. to $\eta$, $V \in C^\infty(N)$ is the \textbf{potential} for $X$, $g$ is a Riemannian metric on $N$ and $\kappa$ a fibre metric on $K$.

We only allow Lie algebras $\mathfrak{g}$ admitting an \textbf{$\mathrm{ad}$-invariant scalar product} to which $\kappa$ shall restrict to on each fibre. Doing so, we achieve infinitesimal gauge invariance for $\mathcal{L}_{\mathrm{YMH}}$, \textit{i.e.}~
\ba
\mleft.\frac{\mathrm{d}}{\mathrm{d}t}\mright|_{t=0}
\mathcal{L}_{\mathrm{YMH}}\mleft[A + t \cdot \delta_\varepsilon A, X\mright]
&=
0
\ea
for $t \in \mathbb{R}$ and all $\varepsilon \in \Gamma(X^*K)$.

We then say that we have a \textbf{Yang-Mills-Higgs gauge theory (YMH GT) for LABs} (with zero minimal coupling to the (Higgs) field $X$). When $\nabla$ is not flat, then we call that as \textbf{curved Yang-Mills-Higgs gauge theory (CYMH GT) for LABs}.

$K$ is a Lie algebroid when equipped with a zero anchor. In the classical setting that would be a gauge theory where the gauge bosons are not coupled to another fields (like the Higgs field) via the minimal coupling. Hence, when one wants to understand $X$ as the Higgs field, then this presents the covariantized Yang-Mills-Higgs theory of massless gauge bosons. In general, $X$ can be any physical field with a Lagrangian similar to the Higgs field, and we will simply refer to $X$ as free physical field.
\end{situations}
\begin{remark}
\leavevmode\newline\label{RemWeDoNotFixZeta}
In the following we want to test whether a given connection $\nabla$ satisfies the compatibility conditions~\eqref{CondSGleichNullLAB} and~\eqref{CondKruemmungmitBLAB}. Especially about the latter we say that a connection $\nabla$ satisfies compatibility condition~\eqref{CondKruemmungmitBLAB} when there is a $\zeta \in \Omega^2(N; K)$ such that this condition is satisfied. So, in general, we are not going to study this condition with respect to a fixed $\zeta$.
\end{remark}

There is a local isomorphism to the classical gauge theory under certain circumstances, keeping the same notation and elements as defined before; this is one of the main motivations of this theory, it generalizes the standard formulation of gauge theory (here of a certain type).
%
%

\begin{corollaries}{Isomorphism to classical gauge theories, \cite[here for LABs]{CurvedYMH}}{CYMHGTGleichClassicalGT}
Let $N$ be a vector space $W$. Assume we have a YMH GT for LABs as in~\ref{sit:CYMHGTForLABsToDoList}, with an LAB $K$ with underlying Lie algebra $\mathfrak{g}$ and equipped with a flat connection $\nabla$. Moreover, $\zeta = 0$.

Then all the formulas of~\ref{sit:CYMHGTForLABsToDoList} are locally the same as for standard Yang-Mills gauge theories for the Lie algebra $\mathfrak{g}$ and zero Lie algebra representation on $W$. 
That means, that we arrive locally at a Yang-Mills gauge theory with an additional free physical field $X$ with potential $V$. When additionally $W = \{*\}$ then we just get a Yang-Mills gauge theory.
\newline

Moreover, each Yang-Mills gauge theory (with or without a free physical field $X$ in a potential $V$) can be described as a YMH GT.
\end{corollaries}

\begin{remark}
\leavevmode\newline\label{WePrueftManDenIsomorphismusZuStandardTheorie}
The idea for the proof is the following: Since $\nabla$ is flat, we have a local parallel frame $\mleft( e_a \mright)_a$, and its pull-back frame $\mleft( X^*e_a \mright)_a$. Expressing everything with respect to this frame results into the same formulas as in classical gauge theories when $\zeta \equiv 0$.

By compatibility condition~\eqref{CondSGleichNullLAB} one can show that the structure functions $C^a_{bc} ~ e_a = \mleft[ e_b, e_c \mright]_{K}$ are constant and then clearly coincide with the ones coming from $\mleft[ X^*e_b, X^*e_c \mright]_{X^*K}$. Therefore the parallel frame spans a Lie algebra $\mathfrak{g}^\prime$ at each point which is isomorphic to $\mathfrak{g}$, and $K$ looks locally like a trivial LAB $U \times \mathfrak{g}^\prime$ with fibre type $\mathfrak{g}^\prime$ ($U$ some open subset). Thus, the idea of the proof can also be described as limiting the space of sections of $K$ locally to constant sections of $U \times \mathfrak{g}^\prime$, and the connection $\nabla$ is then the canonical flat connection.

It is clear that we can construct a YMH GT when starting with a standard Yang-Mills gauge theory, one can just take $K= N \times \mathfrak{g}$, $\zeta \equiv 0$ and its standard flat connection. That the compatibility conditions~\eqref{CondSGleichNullLAB} and~\eqref{CondKruemmungmitBLAB} are satisfied is trivial to check; for the former, observe that it is a tensorial equation, and then test it just with respect to constant sections.
\end{remark}

This corollary motivates the following definitions.

\begin{definitions}{Classical gauge theory}{ClassicalGT}
We say that we have a \textbf{pre-classical gauge theory} when $\nabla$ is flat.

When we have additionally $\zeta = 0$, then we say that we have a \textbf{classical gauge theory}.
\end{definitions}

That is the situation regarding gauge theory and its formalism on Lie algebra bundles. We also have a transformation which keeps the Lagrangian invariant. We will mainly study this transformation in the following sections.

\begin{fieldredefinitions}{Field redefinition in the situation of LABs}{FieldRedefForLABs}
Let us have a (C)YMH GT as in~\ref{sit:CYMHGTForLABsToDoList} and $\lambda \in \Omega^1(N; K)$, then we define the \textbf{field redefinitions} by
\ba\label{TrafoFormelFuerDieEichbosonenA}
\widetilde{A}^\lambda
&\coloneqq
A
	+ \mleft( X^*\lambda \mright)(\mathrm{D}X)
=
A
	+ X^! \lambda, \\
\widetilde{\zeta}^\lambda
&\coloneqq
\zeta
	- \mathrm{d}^\nabla \lambda
	+ \frac{1}{2} \mleft[ \lambda \stackrel{\wedge}{,} \lambda \mright]_K, \label{EqZetaTrafoForLAB}
\ea
and
\ba\label{EqWennFlachDannExaktOderHaltInner}
\widetilde{\nabla}_Y^\lambda \mu
&\coloneqq
\nabla_Y \mu
	- \mleft[ \lambda(Y) , \mu \mright]_K
\ea
for all $Y \in \mathfrak{X}(N)$ and $\mu \in \Gamma(K)$. $X$, the LAB $K$ and the metrics $\kappa$ and $g$ stay the same.

Then the field strength $G$ stays invariant under these transformations, \textit{i.e.}
\ba
\widetilde{G}^\lambda
&\coloneqq
\mathrm{d}^{X^*\widetilde{\nabla}^\lambda}\widetilde{A}^\lambda
	+ \frac{1}{2} \mleft[ \widetilde{A}^\lambda \stackrel{\wedge}{,} \widetilde{A}^\lambda \mright]_{X^*K}
	+ \frac{1}{2} \mleft( X^*\widetilde{\zeta}^\lambda \mright)\mleft( \mathrm{D}X \stackrel{\wedge}{,} \mathrm{D}X \mright)
\equiv
G_{A, X}
\ea 
for all $\lambda$. Thus, the Lagrangian stays invariant, too. $\widetilde{\nabla}^\lambda$ and $\widetilde{\zeta}^\lambda$ satisfy the compatibility conditions, Eq.~\eqref{CondSGleichNullLAB} and~\eqref{CondKruemmungmitBLAB}, and, hence, gauge invariance is preserved, although the infinitesimal gauge transformation changes its shape in Eq.~\eqref{EqInfGaugeTrafoLABs}.
\newline

When we apply this transformation again with $-\lambda$, denoting the corresponding terms with $\widehat{A}^{-\lambda} \coloneqq \widetilde{\widetilde{A}^\lambda}^{-\lambda}$ \textit{etc.}, then we get
\ba
\widehat{A}^{-\lambda} &= A,&
\widehat{\zeta}^{-\lambda} &= \zeta,&
\widehat{\nabla}^{-\lambda} &= \nabla.
\ea
\end{fieldredefinitions}

\begin{remark}
\leavevmode\newline
For Eq.~\eqref{EqWennFlachDannExaktOderHaltInner} we will write 
\ba
\widetilde{\nabla}^\lambda
&=
\nabla
- \mathrm{ad} \circ \lambda,
\ea
where $\mathrm{ad} \circ \lambda \in \Omega^1(N; \mathrm{End}(K))$, $\mleft(\mathrm{ad} \circ \lambda \mright)(Y)(\mu) \coloneqq \mleft[ \lambda(Y), \mu \mright]_K$ for all $Y \in \mathfrak{X}(N)$ and $\mu \in \Gamma(K)$. This implies that 
\bas
(\mathrm{ad} \circ \lambda)(\mu)
&=
\mleft[ \lambda, \mu \mright]_K
=
\mleft[ \lambda \stackrel{\wedge}{,} \mu \mright]_K.
\eas
Similarly, we get $\mathrm{ad} \circ \omega \in \Omega^l(N; \mathrm{End}(K))$.
\end{remark}

Before we prove this, let us discuss it. Using that field redefinition one can motivate the additional term related to $\zeta$ arising in the field strength $G$. Recall Cor.~\ref{cor:CYMHGTGleichClassicalGT}: When $\zeta \equiv 0$, then $\nabla$ is flat by compatibility condition~\eqref{CondKruemmungmitBLAB}, and, hence, we have locally a standard formulation of gauge theory, such that $G = F$, where $F$ is the typical field strength given for Yang-Mills gauge theories. Therefore one could say that $G = F +X^!\zeta$. With the field redefinition one can motivate this term in standard gauge theories.

\begin{remarks}{Another motivation for the new field strength}{MotivationForTheNewFieldStrength}
The idea of the 2-form $\zeta$ in Eq.~\eqref{defNewFieldStrengthG} was that it describes an auxiliary map to allow non-flat $\nabla$ for gauge theories by compatibility condition~\eqref{CondKruemmungmitBLAB}. But let us give a more physical approach how to motivate this 2-form, using the introduced field redefinition and Cor.~\ref{cor:CYMHGTGleichClassicalGT}.

Start with a classical formulation of gauge theory with Lie algebra $\mathfrak{g}$ and extend this to a YMH GT as in~\ref{sit:CYMHGTForLABsToDoList}, \textit{e.g.}~by choosing a trivial Lie algebra bundle $K = N \times \mathfrak{g}$ over $N$, $\zeta = 0$ and taking its canonical flat connection. As explained in \cite{CurvedYMH} and given by Cor.~\ref{cor:CYMHGTGleichClassicalGT}, this theory describes a covariantized formulation of Yang-Mills gauge theories with a free physical field $X$.

Normally, Yang-Mills gauge theory would be with $N = \{*\}$ and then there is only $\zeta \equiv 0$. But one could now add a free physical field $X$ with its target manifold $N$ and change the Lagrangian as in~\eqref{defLagrangianForLABs}. Even when we start with $\zeta = 0$ we could motivate non-zero $\zeta$ in the definition of the Field strength by applying the field redefinition~\ref{fieldredef:FieldRedefForLABs}. Since the field redefinition keeps the Lagrangian invariant we arrive at a Lagrangian with a field strength as given in Eq.~\eqref{defNewFieldStrengthG}, but it still describes the same physics. Although there is no minimal coupling to $X$ of the gauge bosons in the classical setting, the Lagrangian would now give terms where there are products of $A$ with $\mathrm{D}X$. Hence, there is hope for new physics when one allows any $\zeta$ satisfying compatibility condition~\eqref{CondKruemmungmitBLAB} and not just the ones coming from $\zeta \equiv 0$. There is then the natural question whether there are such more general $\zeta$. We try to answer this question here for LABs, but we can already try to give an answer for the abelian situation:

For abelian Lie algebras we would get a $\zeta$ of the form $\widetilde{\zeta}^\lambda = -\mathrm{d}^\nabla \lambda$ when starting with $\zeta \equiv 0$ and then applying the field redefinition. Observe that the connection stays invariant and has to be always flat by compatibility condition~\eqref{CondKruemmungmitBLAB}, thence, $\mathrm{d}^\nabla$ is always a differential. Moreover, compatibility condition~\eqref{CondKruemmungmitBLAB} would be trivial such that there is in general no restriction on $\zeta$. Thence, you can take every $\zeta \in \Omega^2(N; K)$, especially ones which are not exact with respect to $\mathrm{d}^\nabla$ and even ones with $\mathrm{d}^\nabla \zeta \neq 0$ to avoid local exactness. In that way we avoid that there is a field redefinition which could lead to $\widetilde{\zeta}^\lambda = 0$ since this would otherwise imply 
\bas
\mathrm{d}^\nabla \zeta ~ &\stackrel{\text{Eq.~\eqref{EqZetaTrafoForLAB}}}{=} ~ \mleft( \mathrm{d}^\nabla \mright)^2\lambda
~ \stackrel{\nabla \text{ flat}}{=} ~ 0.
\eas
We will come back to this when we have a better understanding of everything, we are also going to give an interpretation of $\mathrm{d}^\nabla \zeta$ at the very end.
\end{remarks}

We need to show several identities for the proof of Thm.~\ref{fieldredef:FieldRedefForLABs} and for some statements about the Bianchi identity of $G$ at the end of this paper, thus, recall \ref{CalculusIdentitiesNeeded}.

\begin{proof}[Proof of Thm.~\ref{fieldredef:FieldRedefForLABs}]
\leavevmode\newline
$\bullet$ We calculate for $A \in \Omega^1(M; X^*K)$
\ba\label{eqWedgeLieklammerbeiPullbackMitA}
X^!\mleft( \mathrm{ad} \circ \lambda \mright) \wedge A
&\stackrel{\text{Eq.~\eqref{EqCommutationRelation}}}{=}
\mleft( \mathrm{ad}^* \circ X^!\lambda \mright) \wedge A
\stackrel{\text{Eq.~\eqref{wedgeproduktmitadLambdaergibtLieklammer}}}{=}
\mleft[ X^!\lambda \stackrel{\wedge}{,} A \mright]_{X^*K}.
\ea
Then observe, using the definition of the field redefinition,
\bas
\widetilde{G}^\lambda
&=
\mathrm{d}^{X^*\widetilde{\nabla}^\lambda}\mleft( A + X^! \lambda \mright)
	+ \frac{1}{2} \mleft[ A + X^! \lambda \stackrel{\wedge}{,} A + X^! \lambda \mright]_{X^*K}
	+ X^!\mleft( \zeta 	- \mathrm{d}^\nabla \lambda 	+ \frac{1}{2} \mleft[ \lambda \stackrel{\wedge}{,} \lambda \mright]_K \mright) \\
&\stackrel{\mathclap{\text{Eq.~\eqref{EqGeilePullBackCommuteFormel}}}}{=}~~~
\mathrm{d}^{X^*\nabla - X^!\mleft( \mathrm{ad} \circ \lambda \mright)} A + X^!\mleft( \mathrm{d}^{\nabla - \mathrm{ad} \circ\lambda} \lambda \mright) \\
&\hspace{1cm}
	+ \frac{1}{2} \mleft[ A \stackrel{\wedge}{,} A \mright]_{X^*K} 
	+ \frac{1}{2} \mleft[ X^! \lambda \stackrel{\wedge}{,} A \mright]_{X^*K} 
	+ \frac{1}{2} \underbrace{\mleft[ A \stackrel{\wedge}{,} X^! \lambda \mright]_{X^*K}}_{\mathclap{\stackrel{\text{Eq.~\eqref{VertauschungsregelForKKlammerAufFormen}}}{=}~ \mleft[ X^! \lambda \stackrel{\wedge}{,} A \mright]_{X^*K}}}
	+ \frac{1}{2} \underbrace{\mleft[ X^! \lambda \stackrel{\wedge}{,} X^! \lambda \mright]_{X^*K}}_{\stackrel{\text{Eq.~\eqref{eqPullbackofLiebracketStuff}}}{=} X^!\mleft( \mleft[ \lambda \stackrel{\wedge}{,} \lambda \mright]_K \mright)} \\
&\hspace{1cm}
	+ X^!\zeta 	- X^! \mleft(\mathrm{d}^\nabla \lambda \mright) 	+ \frac{1}{2} ~ X^! \mleft(\mleft[ \lambda \stackrel{\wedge}{,} \lambda \mright]_K \mright) \\
&\stackrel{\mathclap{\text{Eq.~\eqref{eqDifferentialSplit}}}}{=}~~~
G 
	\underbrace{- X^!\mleft( \mathrm{ad} \circ \lambda \mright) \wedge A
	+ \mleft[ X^! \lambda \stackrel{\wedge}{,} A \mright]_{X^*K}}
	_{\stackrel{\text{Eq.~\eqref{eqWedgeLieklammerbeiPullbackMitA}}}{=} 0}
	\underbrace{- X^!\mleft( \mleft( \mathrm{ad} \circ \lambda \mright) \wedge \lambda
	+ \mleft[ \lambda \stackrel{\wedge}{,} \lambda \mright]_K \mright)}
	_{\stackrel{\text{Eq.~\eqref{wedgeproduktmitadLambdaergibtLieklammer}}}{=} 0} \\
&=
G.
\eas

$\bullet$ Now let us check the compatibility conditions~\eqref{CondSGleichNullLAB} and~\eqref{CondKruemmungmitBLAB},
\bas
\widetilde{\nabla}^\lambda\mleft( \mleft[ \mu, \nu \mright]_K \mright)
&=
\mleft( \nabla - \mathrm{ad}\circ\lambda \mright)\mleft( \mleft[ \mu, \nu \mright]_K \mright) \\
&=
\mleft[ \nabla \mu, \nu \mright]_K + \mleft[ \mu, \nabla \nu \mright]_K
	- \underbrace{\mleft[ \lambda, \mleft[ \mu, \nu \mright]_K \mright]_K}_{\mathclap{= ~ \mleft[ \mu, \mleft[ \lambda, \nu \mright]_K \mright]_K
	+ \mleft[ \mleft[ \lambda, \mu \mright]_K, \nu \mright]_K}} \\
&=
\mleft[ \mleft( \nabla - \mathrm{ad}\circ\lambda \mright) \mu, \nu \mright]_K
	+ \mleft[ \mu, \mleft( \nabla - \mathrm{ad}\circ\lambda \mright) \nu \mright]_K \\
&=
\mleft[ \widetilde{\nabla}^\lambda\mu, \nu \mright]_K
	+ \mleft[ \mu, \widetilde{\nabla}^\lambda \nu \mright]_K
\eas
for all $\mu, \nu \in \Gamma(K)$, using that $\nabla$ satisfies compatibility condition~\eqref{CondSGleichNullLAB} and the Jacobi identity on $\mathfrak{X}(N) \ni Y \mapsto \mleft[ \lambda(Y), \mleft[ \mu, \nu \mright]_K \mright]_K$. For the curvature we calculate
\bas
\widetilde{\nabla}^\lambda_Y \widetilde{\nabla}^\lambda_Z \mu
&=
\Big( \nabla_Y - \mathrm{ad}(\lambda(Y)) \Big)\mleft( \nabla_Z \mu - \mleft[ \lambda(Z), \mu \mright]_K \mright) \\
&=
\nabla_Y \nabla_Z \mu 
	- \mleft[ \nabla_Y \mleft(\lambda(Z)\mright), \mu \mright]_K
	- \mleft[ \lambda(Z), \nabla_Y \mu \mright]_K
	- \mleft[ \lambda(Y), \nabla_Z \mu \mright]_K
	+ \mleft[ \lambda(Y), \mleft[ \lambda(Z), \mu \mright]_K \mright]_K
\eas
for all $Y, Z \in \mathfrak{X}(N)$ and $\mu \in K$, using that $\nabla$ satisfies compatibility condition~\eqref{CondSGleichNullLAB}, and, hence,
\bas
R_{\widetilde{\nabla}^\lambda}(Y, Z)\mu
&=
R_\nabla(Y, Z) \mu
	+ \Big[ \underbrace{- \nabla_Y(\lambda(Z)) + \nabla_Z(\lambda(Y)) + \lambda([Y, Z])}_{= ~- \mathrm{d}^\nabla \lambda}, \mu \Big]_K \\
&\hspace{1cm}
	+ \underbrace{\mleft[ \lambda(Y), \mleft[ \lambda(Z), \mu \mright]_K \mright]_K
	- \mleft[ \lambda(Z), \mleft[ \lambda(Y), \mu \mright]_K \mright]_K}
	_{\mleft[ \mleft[ \lambda(Y), \lambda(Z) \mright]_K, \mu \mright]_K} \\
&=
\mleft[ \mleft(\zeta - \mathrm{d}^\nabla \lambda + \frac{1}{2} \mleft[ \lambda \stackrel{\wedge}{,} \lambda \mright]_K \mright)(Y, Z), \mu  \mright]_K \\
&=
\mleft[ \widetilde{\zeta}^\lambda(Y, Z), \mu \mright]_K
\eas
using that $\nabla$ satisfies compatibility condition~\eqref{CondKruemmungmitBLAB} with $\zeta$. By~\ref{sit:CYMHGTForLABsToDoList} we know that gauge invariance follows since we still get $\delta_\varepsilon G = \mleft[ \varepsilon, G \mright]_{X^*K}$ for all $\varepsilon \in \Gamma(X^*K)$ and since $\kappa$ and $X$ do not transform.

$\bullet$ Now we transform again with $-\lambda$
\bas
\widehat{A}^{-\lambda}
&=
\underbrace{\widetilde{A}^\lambda}_{\mathclap{= ~ A + X^! \lambda}}
	- ~ X^!\lambda
=
A, \\
\widehat{\nabla}^{-\lambda}
&=
\underbrace{\widetilde{\nabla}^{\lambda}}_{\mathclap{= ~\nabla - \mathrm{ad} \circ \lambda}}
	+ ~ \mathrm{ad} \circ \lambda
=
\nabla, \\
\widehat{\zeta}^{-\lambda}
&=
\underbrace{\widetilde{\zeta}^\lambda}_{\mathclap{=~ \zeta - \mathrm{d}^\nabla \lambda + \frac{1}{2} \mleft[ \lambda \stackrel{\wedge}{,} \lambda \mright]_K}}
	+ ~ \mathrm{d}^{\widetilde{\nabla}^\lambda} \lambda
	+ \frac{1}{2} \mleft[ \lambda \stackrel{\wedge}{,} \lambda \mright]_K
\stackrel{\text{Eq.~\eqref{eqDifferentialSplit}}}{=}
\zeta + \mleft[ \lambda \stackrel{\wedge}{,} \lambda \mright]_K
	- (\mathrm{ad} \circ \lambda) \wedge \lambda
\stackrel{\text{Eq.~\eqref{wedgeproduktmitadLambdaergibtLieklammer}}}{=}
\zeta.
\eas
\end{proof}

\section{Relation of vector bundle connections in gauge theories with certain Lie derivation laws} \label{ConnectionIsALieDerivation}
Starting with a CYMH GT for a given LAB $K \to N$, by Cor.~\ref{cor:CYMHGTGleichClassicalGT} there is the natural question whether or not one arrives at a (pre-)classical gauge theory by using the field redefinition~\ref{fieldredef:FieldRedefForLABs}.

In order to do so it is important to understand what type of connection $\nabla$ we have due to the compatibility conditions~\eqref{CondSGleichNullLAB} and~\eqref{CondKruemmungmitBLAB}. Let us look at a general connection $\nabla$ in a slightly different way, as also introduced in \cite[\S 5.2, the discussion after Def. 5.2.3; page 185]{mackenzieGeneralTheory}: By the very definition of a vector bundle connection we have on one hand
\bas
\nabla_Y (\alpha \mu + \beta \nu)
&=
\alpha \nabla_Y \mu + \beta \nabla_Y \nu, \\
\nabla_Y (f \mu)
&=
f \nabla_Y \mu + Y(f) ~ \mu
\eas
for all $Y \in \mathfrak{X}(N)$, $\mu, \nu \in \Gamma(K)$, $\alpha, \beta \in \mathbb{R}$ and $f \in C^\infty(N)$. That is equivalent to say that $\nabla_Y$ is a derivation of the vector bundle $K$ for all $Y \in \mathfrak{X}(N)$:

\begin{definitions}{Derivations on a vector bundle $V$, \newline \cite[\S 3.3, Example 3.3.4; page 102f.]{mackenzieGeneralTheory}}{DerivationsOnV}
Let $V \to N$ be a real vector bundle with finite rank. Then a \textbf{derivation on $V$} is an $\mathbb{R}$-linear map $\mathcal{T}: \Gamma(V) \to \Gamma(V)$ such that there is a smooth vector field $a\mleft( \mathcal{T} \mright) \in \mathfrak{X}(N)$ with
\ba\label{eqDerivationsLiftASuperDuperVectorField}
\mathcal{T}(fv)
&=
f ~ \mathcal{T}(v)
	+ a\mleft( \mathcal{T} \mright)(f) ~ v
\ea
for all $f \in C^\infty(N)$ and $v \in \Gamma(V)$. We say that $\mathcal{T}$ lifts $a(\mathcal{T})$.
\end{definitions}

As argued in \cite[\S 3.3, Example 3.3.4; page 102f.]{mackenzieGeneralTheory}, derivations of $K$ are sections of a transitive\footnote{That means, its anchor is surjective.} Lie algebroid $\mathcal{D}(K)$, where the bracket is given by the commutator of derivations and the surjective anchor by $a$; especially, the endomorphisms of $K$ form precisely the kernel of $a$, which is also a Lie subalgebroid of $\mathcal{D}(K)$.

On the other hand, we also have
\bas
\nabla_{fY+hZ}
&=
f\nabla_Y+h\nabla_Z
\eas
for all $Y, Z \in \mathfrak{X}(N)$ and $f, h \in C^\infty(N)$. By the Leibniz rule we also have
\bas
a(\nabla_Y) &= Y.
\eas
This finally implies that $\nabla$ comes from an anchor-preserving\footnote{The tangent bundle is equipped with the identity anchor as usual.} (and base-preserving) vector bundle morphism, or, equivalently, a morphism of anchored vector bundles, $\gamma: \mathrm{T}N \to \mathcal{D}(K)$, that is
\bas
a(\gamma(Y)) &= Y, \\
\gamma(Y)(\mu)
&=
\nabla_Y \mu
\eas
for all $Y \in \mathfrak{X}(N)$ and $\mu \in \Gamma(K)$. We are going to denote both interpretations by $\nabla$. Thus, there is a 1:1 correspondence of vector bundle connections with anchor-preserving vector bundle morphisms from $\mathrm{T}N$ to $\mathcal{D}(K)$. With that notion it is easy to check that $\nabla$ is flat as a connection if and only if $\nabla$ is a Lie algebroid morphism, see \textit{e.g.}~\cite[\S 5.2, Definition 5.2.9; page 187]{mackenzieGeneralTheory}.

Now back to the compatibility conditions. For all $Y \in \mathfrak{X}(N)$, compatibility condition~\eqref{CondSGleichNullLAB} implies that $\nabla_Y$ is a derivation of the Lie bracket $\mleft[ \cdot, \cdot \mright]_K$ and so of $\mleft[ \cdot, \cdot \mright]_{\mathfrak{g}}$ on each fibre. Thence, the vector bundle morphism $\nabla$ has values in $\mathcal{D}_{\mathrm{Der}}(K)$, the subbundle of $\mathcal{D}(K)$ of derivations which are also derivations of the Lie bracket.

$\mathcal{D}_{\mathrm{Der}}(K)$ is also a (transitive) Lie subalgebroid of $\mathcal{D}(K)$ as discussed in \cite[\S 3.6, discussion around Corollary 3.6.11; page 140f.]{mackenzieGeneralTheory}. So, we arrive at that $\nabla$ has to be a so-called Lie derivation law by compatibility condition~\eqref{CondSGleichNullLAB}:

\begin{definitions}{Lie derivation law, \newline \cite[\S 7.2, Definition 7.2.9, page 275]{mackenzieGeneralTheory}}{LieConnection}
Let $K \to N$ be an LAB. A \textbf{Lie derivation law} for $\mathrm{T}N$ with coefficients in $K$ is an anchor- and base-preserving vector bundle morphism $\nabla: \mathrm{T}N \to \mathcal{D}_{\mathrm{Der}}(K)$, that is, a connection $\nabla$ on $K$ in the usual sense such that
\ba
\nabla_Y\mleft( \mleft[ \mu, \nu \mright]_K \mright)
&=
\mleft[ \nabla_Y \mu, \nu \mright]_K
	+ \mleft[ \mu, \nabla_Y \nu \mright]_K
\ea
for all $Y \in \mathfrak{X}(N)$ and $\mu, \nu \in \Gamma(K)$.
\end{definitions}

\begin{remark}
\leavevmode\newline
Consider a bundle of Lie algebras of the same rank (which is not necessarily an LAB), then the existence of a Lie derivation law for $\mathrm{T}N$ with coefficients in $K$ is also equivalent to say that it is an LAB. Thence, it is not necessary to look at bundle of Lie algebras in this context of gauge theory by compatibility condition~\eqref{CondSGleichNullLAB}, see \textit{e.g.}~\cite[\S 6.4, Theorem 6.4.5; page 238]{mackenzieGeneralTheory} and \cite[\S 2, Proposition 2.13]{basicconn}.

As argued in \cite[\S 5.2, second part of Example 5.2.12 where it is also called a Lie connection; page 188f.]{mackenzieGeneralTheory} such a Lie derivation law always exists for a given LAB.
\end{remark}

The kernel of $a|_{\mathcal{D}_{\mathrm{Der}}(K)}$ is exactly the sub-LAB $\mathrm{Der}(K)$, derivations of $K$ with fibre type $\mathrm{Der}(\mathfrak{g})$, the Lie bracket derivations of $\mathfrak{g}$ (\textit{i.e.}~$\mathrm{Der}(K)$ consists of Lie bracket derivations which are also endomorphisms of $K$; see \cite[\S 3.3; page 104ff.]{mackenzieGeneralTheory} for its construction)
\begin{center}
	\begin{tikzcd}
	\mathrm{Der}(\mathfrak{g}) \arrow{r} & \mathrm{Der}(K) \arrow{d} \\
			& N
	\end{tikzcd}
\end{center}
All of this results then into the following short exact sequence of Lie algebroids\footnote{The hooked arrow emphasizes the inclusion, and the two-headed arrow the surjectivity.}
\be\label{ShortSequenceOfDerivationsWithBracketderivation}
	\begin{tikzcd}
		\mathrm{Der}(K) \arrow[r, hook] & \mathcal{D}_{\mathrm{Der}}(K) \arrow[two heads]{r}{a} & \mathrm{T}N.
	\end{tikzcd}
\ee

Now about understanding the compatibility condition~\eqref{CondKruemmungmitBLAB}: In the context of the field redefinition, when it would be possible to make $\nabla$ flat by a field redefinition then there would be a parallel frame $\mleft( e_a \mright)_a$ locally for $\widetilde{\nabla}^\lambda$ such that by Eq.~\eqref{EqWennFlachDannExaktOderHaltInner}
\bas
\nabla_Y e_a
&=
\mleft[ \lambda(Y), e_a \mright]_K
\eas
for all $Y \in \mathfrak{X}(N)$. That is, with respect to that frame, the Lie bracket derivation $\nabla_Y$ looks like an adjoint of $\lambda(Y)$, an inner Lie bracket derivation. Thence, it makes sense to look at the Lie algebroid of outer derivations \textit{etc.}. For that recall that one can define other terms for $K$ similar to the terms for the Lie algebra $\mathfrak{g}$, see \cite[\S 3.3; page 104ff.]{mackenzieGeneralTheory} for the following definitions and their relations: The LABs $Z(K)$ (center of K with fibre type $Z(\mathfrak{g})$, the centre of $\mathfrak{g}$)
\begin{center}
	\begin{tikzcd}
	Z(\mathfrak{g}) \arrow{r}	& Z(K) \arrow{d} \\
			& N
	\end{tikzcd}
\end{center}
and $\mathrm{ad}(K)$ (the adjoint of $K$ with fibre type $\mathrm{ad}(\mathfrak{g})$, the adjoint of $\mathfrak{g}$)
\begin{center}
	\begin{tikzcd}
	\mathrm{ad}(\mathfrak{g}) \arrow{r} & \mathrm{ad}(K) \arrow{d} \\
			& N
	\end{tikzcd}
\end{center}
$\mathrm{ad}(K)$ is an ideal of the LABs $\mathrm{Der}(K)$ and $\mathcal{D}_{\mathrm{Der}}(K)$ and we can define their quotient as Lie algebroids over $\mathrm{ad}(K)$, see \cite[\S 7.2, Definition 7.2.1; page 271f.]{mackenzieGeneralTheory}:~\eqref{ShortSequenceOfDerivationsWithBracketderivation} results into a new short exact sequence of Lie algebroids
\be\label{ShortSequenceOfDerivationsWithBracketderivationAndQuotient}
	\begin{tikzcd}
		\mathrm{Der}(K)\Big/\mathrm{ad}(K) \arrow[r, hook] & \mathcal{D}_{\mathrm{Der}}(K)\Big/ \mathrm{ad}(K) \arrow[two heads]{r}{\overline{a}} & \mathrm{T}N,
	\end{tikzcd}
\ee
which is denoted by
\be\label{defShortSequenceOfDerivationsWithBracketderivationAndQuotient}
	\begin{tikzcd}
		\mathrm{Out}(K) \arrow[r, hook] & \mathrm{Out}\mleft(\mathcal{D}_{\mathrm{Der}}(K)\mright) \arrow[two heads]{r}{\overline{a}} & \mathrm{T}N,
	\end{tikzcd}
\ee
and we call $\mathrm{Out}\mleft(\mathcal{D}_{\mathrm{Der}}(K)\mright)$ the Lie algebroid of \textbf{outer bracket derivations of $K$} with anchor $\overline{a}$, canonically induced by $a$. In total we arrive at the following commuting diagram
\be\label{theFullDiagramForLABStuff}
	\begin{tikzcd}
		Z(K) \arrow[hook]{d}{i} \arrow[equal]{r} & Z(K) \arrow[hook]{d} \\
		K \arrow{d}{\mathrm{ad}} \arrow[equal]{r} & K \arrow{d} \\
		\mathrm{Der}(K) \arrow[two heads]{d}{\sharp^+} \arrow[hook]{r}{j} & \mathcal{D}_{\mathrm{Der}}(K) \arrow[two heads]{d}{\sharp} \arrow[two heads]{r}{a} & \mathrm{T}N \arrow[equal]{d} \\
		\mathrm{Out}(K) \arrow[hook]{r}{\overline{j}} & \mathrm{Out}\mleft(\mathcal{D}_{\mathrm{Der}}(K)\mright) \arrow[two heads]{r}{\overline{a}} & \mathrm{T}N
	\end{tikzcd}
\ee
as introduced in \cite[\S 7.2, Figure 7.1; page 272]{mackenzieGeneralTheory}, where both rows and columns are short exact sequences of Lie algebroids and the diagram serves as a definition of the notation of the new Lie algebroid morphisms, especially, $\sharp$ denotes the projection of derivations into the space of outer derivations. 

A note about the notation: $i$, $j$ and $\overline{j}$ are for example just the inclusions when using the standard descriptions of these Lie algebroids. But these notations allow a change of the explicit description of these Lie algebroids, in that case the inclusions would be replaced by a composition of the corresponding inclusions with a Lie algebroid isomorphism. In our case we use the standard definitions such that $i$, $j$ and $\overline{j}$ are inclusions, hence, we will omit them in the following.

With diagram~\eqref{theFullDiagramForLABStuff} we can now finally study compatibility condition~\eqref{CondKruemmungmitBLAB}. The curvature $R_\nabla$ of a Lie connection $\nabla:\mathrm{T}N \to \mathcal{D}_{\mathrm{Der}}(K)$ is clearly an element of $\Omega^2\mleft(N; \mathcal{D}_{\mathrm{Der}}(K)\mright)$ since
\bas
R_\nabla(Y, Z)
&=
\underbrace{\mleft[ \nabla_Y, \nabla_Z \mright]_{\mathcal{D}_{\mathrm{Der}}(K)}}_{\in ~ \Gamma(\mathcal{D}_{\mathrm{Der}}(K))}
	- \underbrace{\nabla_{[Y, Z]}}_{\mathclap{\in ~ \Gamma(\mathcal{D}_{\mathrm{Der}}(K))}}
	\in \Gamma(\mathcal{D}_{\mathrm{Der}}(K))
\eas
for all $Y, Z \in \mathfrak{X}(N)$. Compatibility condition~\eqref{CondKruemmungmitBLAB} is then equivalent to\footnote{Recall Remark~\ref{RemWeDoNotFixZeta}; we understand this compatibility condition now as "there exists a $\zeta$ such that [...]".}
\ba
\sharp\mleft( R_\nabla(Y, Z) \mright) = 0
\ea
for all $Y, Z \in \mathfrak{X}(N)$. We will show that this implies that $\nabla$ is a so-called Lie derivation law covering a pairing of $\mathrm{T}N$ with $K$. For that we need to define what a pairing is.\footnote{Mackenzie called the following construction a \textbf{coupling} and not \textbf{pairing}. I renamed it to avoid confusion with couplings in a physical context}

\begin{definitions}{Pairing of $\mathrm{T}N$, \cite[\S 7.2, Definitions 7.2.2; page 272]{mackenzieGeneralTheory}}{CouplingsOfTNWithK}
A \textbf{pairing} of $\mathrm{T}N$ is a pair of an LAB $K \to N$ together with a (base-preserving) morphism of Lie algebroids $\Xi: \mathrm{T}N \to \mathrm{Out}(\mathcal{D}_{\mathrm{Der}}(K))$. We also say that $\mathrm{T}N$ and $K$ are \textbf{paired by $\Xi$}.
\end{definitions}

Now we can define a special type of connection.

\begin{definitions}{Lie derivation law covering $\Xi$, \newline\cite[\S 7.2, see discussion after Definition 7.2.2; page 272]{mackenzieGeneralTheory}}{LieDerivationLawOverACouplingXi}
Let $K \to N$ be an LAB and $\nabla: \mathrm{T}N \to \mathcal{D}_{\mathrm{Der}}(K)$ a Lie derivation law. Assume that $\mathrm{T}N$ and $K$ are coupled by a (base-preserving) Lie algebroid morphism $\Xi: \mathrm{T}N \to \mathrm{Out}(\mathcal{D}_{\mathrm{Der}}(K))$.
Then we say that $\nabla$ is a \textbf{Lie derivation law covering $\Xi$} if
\ba
\sharp \circ \nabla
&=
\Xi.
\ea
\end{definitions}

\begin{remark}
\leavevmode\newline
So, while a Lie derivation law is not necessarily a morphism of Lie algebroids, $\sharp \circ \nabla$ is of that type when $\nabla$ covers a pairing.
\end{remark}

This type of connection is exactly the type we need for gauge theory on LABs.

\begin{theorems}{(C)YMH GT only allows Lie derivation laws covering $\Xi$}{GaugeTheoryNeedsLieDerivLawsCoveringACoupling}
Let $K \to N$ be an LAB. Then a map $\nabla: \mathrm{T}N \to \mathcal{D}_{\mathrm{Der}}(K)$ is a Lie derivation law covering some (base-preserving) Lie algebroid morphism $\Xi: \mathrm{T}N \to \mathrm{Out}(\mathcal{D}_{\mathrm{Der}}(K))$ if and only if it is a connection on $K$ satisfying the compatibility conditions~\eqref{CondSGleichNullLAB} and~\eqref{CondKruemmungmitBLAB}, \textit{i.e.}
\bas
\nabla_Y\mleft( \mleft[ \mu, \nu \mright]_K \mright)
&=
\mleft[ \nabla_Y \mu, \nu \mright]_K
	+ \mleft[ \mu, \nabla_Y \nu \mright]_K, \\
\sharp\mleft(R_\nabla(Y, Z)\mright)
&=
0
\eas
for all $Y, Z \in \mathfrak{X}(N)$ and $\mu, \nu \in \Gamma(K)$.
\end{theorems}

\begin{remark}\label{remExistenceOfLieDerivationLawsCoveringACoupling}
\leavevmode\newline
So, we have seen that compatibility condition~\eqref{CondSGleichNullLAB} implies that $\nabla$ has to be a Lie derivation law, and compatibility condition~\eqref{CondKruemmungmitBLAB} then implies that it covers a pairing of $\mathrm{T}N$ and $K$.

As argued in \cite[\S 7.2, discussion after Definition 7.2.2, replace the $A$ there with $\mathrm{T}N$; page 272]{mackenzieGeneralTheory}, for a given $\Xi$ there is always a Lie derivation law covering it.
\end{remark}

\begin{proof}
\leavevmode\newline
We already have seen that a connection $\nabla$ satisfying compatibility condition~\eqref{CondSGleichNullLAB} has a 1:1 correspondence to an anchor-preserving vector bundle morphism $\nabla: \mathrm{T}N \to \mathcal{D}_{\mathrm{Der}}(K)$, \textit{i.e.}~a Lie derivation law. So, we only have to care about compatibility condition~\eqref{CondKruemmungmitBLAB}.

"$\Leftarrow$": So, let us have a Lie derivation law with $\sharp\mleft(R_\nabla(Y, Z)\mright) = 0$ for all $Y, Z \in \mathfrak{X}(N)$. Define $\Xi \coloneqq \sharp \circ \nabla$, and recall that $\sharp: \mathcal{D}_{\mathrm{Der}}(K) \to \mathrm{Out}(\mathcal{D}_{\mathrm{Der}}(K))$ is a Lie algebroid morphism such that $\Xi$ is an anchor-preserving vector bundle morphism by definition, using that $\nabla$ is a Lie derivation law,
\bas
\overline{a} \circ \Xi
&=
\overline{a} \circ \sharp \circ \nabla
=
a \circ \nabla
= \mathds{1}_{\mathrm{T}N}.
\eas
Using that $\sharp$ is a homormorphism of Lie brackets, and by $\sharp\mleft(R_\nabla(Y, Z)\mright) = 0$ for all $Y, Z \in \mathfrak{X}(N)$, we also get
\bas
\Xi([Y, Z])
&=
\sharp \mleft( \nabla_{[Y, Z]} \mright)
=
\sharp \mleft( \mleft[ \nabla_Y, \nabla_Z \mright]_{\mathcal{D}_{\mathrm{Der}}(K)} \mright) \\
&=
\mleft[ \sharp\mleft(\nabla_Y\mright), \sharp\mleft(\nabla_Z\mright) \mright]_{\mathrm{Out}(\mathcal{D}_{\mathrm{Der}}(K))}
=
\mleft[ \Xi(Y), \Xi(Z) \mright]_{\mathrm{Out}(\mathcal{D}_{\mathrm{Der}}(K))},
\eas
\textit{i.e.}~$\Xi$ is a Lie algebroid morphism (base-preserving), and it is covered by $\nabla$ due to its definition.

"$\Rightarrow$": This part of the proof is as in \cite[\S 7.2, discussion after Definition 7.2.2; page 272]{mackenzieGeneralTheory} and similar to the previous calculation. Let $\nabla$ be a Lie derivation law covering some Lie algebroid morphism $\Xi$, especially, $\sharp \circ \nabla = \Xi$. That implies
\bas
\sharp\mleft( R_\nabla(Y, Z) \mright)
&=
\sharp\mleft( \mleft[ \nabla_Y, \nabla_Z \mright]_{\mathcal{D}_{\mathrm{Der}}(K)} - \nabla_{[Y, Z]} \mright) \\
&=
\mleft[ \sharp(\nabla_Y), \sharp(\nabla_Z) \mright]_{\mathrm{Out}(\mathcal{D}_{\mathrm{Der}}(K))} - \sharp\mleft(\nabla_{[Y, Z]}\mright) \\
&=
\mleft[ \Xi(Y), \Xi(Z) \mright]_{\mathrm{Out}(\mathcal{D}_{\mathrm{Der}}(K))} - \Xi([Y, Z]) \\
&=
0
\eas
for all $Y, Z \in \mathfrak{X}(N)$, using that both, $\sharp$ and $\Xi$, are homomorphisms of the corresponding Lie brackets. This finishes the proof.
\end{proof}

Given a Lie derivation law covering some $\Xi$, we get that $\nabla$ is an anchor-preserving vector bundle morphism and $\sharp \circ \nabla = \Xi$ is a Lie algebroid morphism. When we want that $\nabla$ is not flat, in the hope of finding a new gauge theory (recall Cor.~\ref{cor:CYMHGTGleichClassicalGT}), we do not want that $\nabla$ itself is a Lie algebroid morphism, while $\sharp$ is a Lie algebroid morphism and $\Xi = \sharp \circ \nabla$, too. That looks like a tightrope walk. Indeed, the field redefinition~\ref{fieldredef:FieldRedefForLABs} may lead to a flat connection while keeping the same physics, \textit{i.e.}~the Lagrangian stays the same.

%

To study this, we now need to construct an invariant for the field redefinition. We will see that this is exactly the so-called \textit{obstruction class} of Mackenzie constructed in \cite[\S 7.2; page 271ff.]{mackenzieGeneralTheory}. 
Observe the following, using the notation as introduced in~\eqref{theFullDiagramForLABStuff}.

\begin{propositions}{Field redefinition preserves the pairing}{FieldRedefPreservesCoupling}
Let $(K, \Xi)$ be a pairing of $\mathrm{T}N$, $\nabla$ be a Lie derivation law covering $\Xi$ and $\zeta \in \Omega^2(N;K)$ satisfying compatibility condition~\eqref{CondKruemmungmitBLAB} with respect to $\nabla$.

Then the field redefinition~\ref{fieldredef:FieldRedefForLABs} preserves the pairing, \textit{i.e.}~$\widetilde{\nabla}^\lambda$ is also a Lie derivation law covering $\Xi$ for all $\lambda \in \Omega^1(N;K)$. Moreover, for every other Lie derivation law $\nabla^\prime$ covering $\Xi$ there is a $\lambda \in \Omega^1(N; K)$ such that
\bas
\nabla^\prime &= \widetilde{\nabla}^\lambda,
\eas
and for its curvature
\bas
R_{\nabla^\prime} &= \mathrm{ad} \circ \widetilde{\zeta}^\lambda.
\eas
\end{propositions}

\begin{remark}
\leavevmode\newline
These are exactly the same formulas as in \cite[\S 7.2, Proposition 7.2.7, identifying Mackenzie's 1-form $l$ with $- \lambda$, also beware that Mackenzie defines curvatures with an opposite sign; page 274]{mackenzieGeneralTheory}. In this reference Mackenzie studies the form given by the difference of two Lie derivation laws covering the same pairing and arrives exactly at our formulas of the field redefinition which we have derived from a physical context of gauge theory on LABs.

In the following we will show that Mackenzie's study has roughly a 1:1 correspondence to the question whether one can find a field redefinition such that $\widetilde{\nabla}^\lambda$ is flat. We have now seen that the field redefinitions just describe a change to any other Lie derivation law covering the same $\Xi$.
\end{remark}

\begin{proof}
\leavevmode\newline
We know that the field redefinition preserves the compatibility conditions~\eqref{CondSGleichNullLAB} and~\eqref{CondKruemmungmitBLAB}, \textit{i.e.}
\bas
\widetilde{\nabla}^\lambda_Y\mleft( \mleft[ \mu, \nu \mright]_K \mright)
&=
\mleft[ \widetilde{\nabla}^\lambda_Y \mu, \nu \mright]_K
	+ \mleft[ \mu, \widetilde{\nabla}^\lambda_Y \nu \mright]_K, \\
R_{\widetilde{\nabla}^\lambda}(Y, Z) \mu
&=
\mleft[ \widetilde{\zeta}^\lambda(Y, Z), \mu \mright]_K,
\eas
that implies by Thm.~\ref{thm:GaugeTheoryNeedsLieDerivLawsCoveringACoupling} that $\widetilde{\nabla}^\lambda$ is a Lie derivation law covering $\widetilde{\Xi}^\lambda \coloneqq \sharp \circ \widetilde{\nabla}^\lambda$. Moreover, using the notation of~\eqref{theFullDiagramForLABStuff},
\bas
\sharp \circ \widetilde{\nabla}^\lambda
&=
\sharp \circ \mleft( \nabla - \mathrm{ad} \circ \lambda \mright)
=
\sharp \circ \nabla
=
\Xi
\eas
for all $\lambda \in \Omega^1(N; K)$, using $\sharp \circ \mathrm{ad} = 0$. This shows that $\widetilde{\nabla}^\lambda$ covers $\Xi$.

Now let $\nabla^\prime$ be another Lie derivation law covering $\Xi$, then clearly
\bas
a|_{\mathcal{D}_{\mathrm{Der}}(K)}(\nabla^\prime_Y - \nabla_Y)
&= Y- Y = 0
\eas
for all $Y \in \mathfrak{X}(N)$, such that $\nabla^\prime - \nabla \in \Omega^1(N; \mathrm{Der}(K))$ by~\eqref{theFullDiagramForLABStuff}, and
\bas
0
&=
\Xi - \Xi
=
\sharp \circ \nabla^\prime
	- \sharp \circ \nabla
=
\sharp \circ \underbrace{\mleft( \nabla^\prime - \nabla \mright)}_{\mathclap{\in ~ \Omega^1(N; \mathrm{Der}(K))}}
=
\sharp^+ \circ \mleft( \nabla^\prime - \nabla \mright).
\eas
Again by~\eqref{theFullDiagramForLABStuff}, there is a $\mu(Y) \in \Gamma(K)$ such that $\nabla^\prime_Y - \nabla_Y = \mathrm{ad}(\mu(Y))$ for all $Y \in \mathfrak{X}(N)$, and due to the $C^\infty$-linearity w.r.t. $Y$ we get $\nabla^\prime - \nabla = \mathrm{ad} \circ \mu$ for a $\mu \in \Omega^1(N; K)$. By field redefinition~\ref{fieldredef:FieldRedefForLABs} we can take $\lambda = - \mu$ to get $\nabla^\prime = \widetilde{\nabla}^\lambda$.

Since $\nabla$ satisfies compatibility condition~\eqref{CondKruemmungmitBLAB} by Thm.~\ref{thm:GaugeTheoryNeedsLieDerivLawsCoveringACoupling} and since this condition is preserved by a field redefinition, the last statement follows, $R_{\nabla^\prime}(Y, Z) = \mathrm{ad}\mleft( \widetilde{\zeta}^\lambda(Y, Z) \mright)$ for all $Y, Z \in \mathfrak{X}(N)$.
\end{proof}


%

Locally we can say the following.

\begin{corollaries}{Local existence of a flat Lie derivation law covering a pairing}{CorLocalerFlacherZusammenhangFuerIrgendeineKopplung}
Let $K$ be an LAB. Then locally there is always a flat Lie derivation law covering some (base-preserving) Lie algebroid morphism $\Xi: \mathrm{T}N \to \mathrm{Out}(\mathcal{D}_{\mathrm{Der}}(K))$.
\end{corollaries}

\begin{remark}
\leavevmode\newline
So, locally, by using Prop.~\ref{prop:FieldRedefPreservesCoupling}, the question whether or not one can transform to a flat connection with the field redefinition breaks down to the question if there is a flat connection covering the same pairing.
\end{remark}

\begin{proof}
\leavevmode\newline
Locally there is a trivialization $K \cong U \times \mathfrak{g}$ as LABs on some open subset $U \subset N$. Then define $\nabla$ as the canonical flat connection, so, the parallel frame is given by constant sections. As in the third paragraph of Remark~\ref{WePrueftManDenIsomorphismusZuStandardTheorie}, the compatibility conditions~\eqref{CondSGleichNullLAB} and~\eqref{CondKruemmungmitBLAB} are satisfied for the canonical flat connection. By Thm.~\ref{thm:GaugeTheoryNeedsLieDerivLawsCoveringACoupling} the statement then follows.
\end{proof}

\section{Obstruction for non-pre-classical gauge theories}\label{MackenzieZeugsUndExistenzvonPreclassical}

\subsection{Obstruction class}

In order to study the field redefinitions~\ref{fieldredef:FieldRedefForLABs}, we now introduce an invariant.

\begin{theorems}{Invariant of the field redefinition}{BianchiIdentityForZeta}
Let $(K, \Xi)$ be a pairing of $\mathrm{T}N$ and $\nabla$ be a Lie derivation law covering $\Xi$. Also let $\zeta$ be any element of  $\Omega^2(N; K)$ that satisfies compatibility condition~\eqref{CondKruemmungmitBLAB} with respect to $\nabla$.

Then we have 
\ba\label{dnablazetaIsCentrevalued}
\mathrm{d}^\nabla \zeta &\in \Omega^3(N; Z(K)),
\ea
\textit{i.e.}~$\mathrm{d}^\nabla \zeta$ has always values in the centre of $K$, $\mathrm{ad} \circ \mathrm{d}^\nabla \zeta = 0$. Moreover,
\ba\label{InvarianteFuerFieldRedefImFallLAB}
\mathrm{d}^{\widetilde{\nabla}^\lambda} \widetilde{\zeta}^\lambda
&=
\mathrm{d}^\nabla \zeta
\ea
for all $\lambda \in \Omega^1(N;K)$ as in~\ref{fieldredef:FieldRedefForLABs}.
\end{theorems}

\begin{remark}
\leavevmode\newline
Eq.~\eqref{dnablazetaIsCentrevalued} and~\eqref{InvarianteFuerFieldRedefImFallLAB} got also derived in a completely different context, see \cite[\S 7.2, Lemma 7.2.4, $\zeta$ is denoted as $\Lambda$ there; page 273]{mackenzieGeneralTheory} and \cite[\S 7.2, Proposition 7.2.11, last statement, there $\zeta$ is denoted by $\Lambda$ and $\mathrm{d}^\nabla \zeta$ by $f(\nabla, \Lambda)$; page 276]{mackenzieGeneralTheory}, respectively; 
see also the discussion about Mackenzie's results later, to know from which context Mackenzie derived these two equations.
\end{remark}

\begin{proof}
\leavevmode\newline
Due to Thm.~\ref{thm:GaugeTheoryNeedsLieDerivLawsCoveringACoupling} we know that we can use the compatibility conditions~\eqref{CondSGleichNullLAB} and~\eqref{CondKruemmungmitBLAB}.

$\bullet$ As also argued in \cite[\S 7.2, Lemma 7.2.4, $\zeta$ is denoted as $\Lambda$ there; page 273]{mackenzieGeneralTheory}, use the well-known Bianchi identity for curvatures $R_\nabla$, \textit{i.e.}
\bas
\mathrm{d}^\nabla R_\nabla = 0,
\eas
viewing $R_\nabla$ as an element of $\Omega^2(N; \mathrm{End}(K))$, and canonically extending $\nabla$ to $\mathrm{End}(K)$, which we still denote by $\nabla$. Compatibility condition~\eqref{CondKruemmungmitBLAB} implies $R_\nabla = \mathrm{ad} \circ \zeta$, and by compatibility condition~\eqref{CondSGleichNullLAB} we can use Eq.~\eqref{DifferentialvonNabalVertauschmitAd}, such that the Bianchi identity implies
\bas
0
&\stackrel{\text{Eq.~\eqref{CondKruemmungmitBLAB}}}{=}
\mathrm{d}^\nabla\mleft( \mathrm{ad} \circ \zeta \mright)
\stackrel{\text{Eq.~\eqref{DifferentialvonNabalVertauschmitAd}}}{=}
\mathrm{ad} \circ \mathrm{d}^\nabla \zeta.
\eas

$\bullet$ Finally, recall that in general curvatures satisfy
\bas
\mleft( \mathrm{d}^\nabla \mright)^2 \omega = R_\nabla \wedge \omega
\eas
for all $\omega \in \Omega^l(N;K)$, viewing $R_\nabla$ as an element of $\Omega^2(N; \mathrm{End}(K))$. Then we have
\bas
\mleft( \mathrm{d}^\nabla \mright)^2 \lambda
&=
R_\nabla \wedge \lambda
\stackrel{\text{Eq.~\eqref{CondKruemmungmitBLAB}}}{=}
(\mathrm{ad} \circ \zeta) \wedge \lambda
\stackrel{\text{Eq.~\eqref{wedgeproduktmitadLambdaergibtLieklammer}}}{=}
\mleft[ \zeta \stackrel{\wedge}{,} \lambda \mright]_K, \\
\mathrm{d}^\nabla \mleft( \mleft[ \lambda \stackrel{\wedge}{,} \lambda \mright]_K \mright)
~~~~&\stackrel{\mathclap{\text{Eq.~\eqref{eqDerivationOfDifferentialOnBracketonK}}}}{=}~~~~
\mleft[ \mathrm{d}^\nabla \lambda \stackrel{\wedge}{,} \lambda \mright]_K
	- \mleft[ \lambda \stackrel{\wedge}{,} \mathrm{d}^\nabla \lambda \mright]_K
\stackrel{\text{Eq.~\eqref{VertauschungsregelForKKlammerAufFormen}}}{=}
2 ~ \mleft[ \mathrm{d}^\nabla \lambda \stackrel{\wedge}{,} \lambda \mright]_K, \\
( \mathrm{ad} \circ \lambda ) \wedge \widetilde{\zeta}^\lambda
~~~~&\stackrel{\mathclap{\text{Eq.~\eqref{wedgeproduktmitadLambdaergibtLieklammer}}}}{=}~~~~
\mleft[ \lambda \stackrel{\wedge}{,} \widetilde{\zeta}^\lambda \mright]_K
\stackrel{\text{Eq.~\eqref{VertauschungsregelForKKlammerAufFormen}}}{=}
- \mleft[ \widetilde{\zeta}^\lambda \stackrel{\wedge}{,} \lambda \mright]_K
\stackrel{\text{Eq.~\eqref{EqZetaTrafoForLAB},~\eqref{JacobiIdentityForFormBracket}}}{=}
- \mleft[ \zeta \stackrel{\wedge}{,} \lambda \mright]_K
	+ \mleft[ \mathrm{d}^\nabla \lambda \stackrel{\wedge}{,} \lambda \mright]_K,
\eas
and, by combining everything, we arrive at
\bas
\mathrm{d}^{\widetilde{\nabla}^\lambda} \widetilde{\zeta}^\lambda
&=
\mathrm{d}^{\nabla - \mathrm{ad} \circ \lambda} \mleft( \widetilde{\zeta}^\lambda \mright)
\stackrel{\text{Eq.~\eqref{eqDifferentialSplit},~\eqref{EqZetaTrafoForLAB}}}{=}
\mathrm{d}^\nabla \mleft( \zeta
	- \mathrm{d}^\nabla \lambda
	+ \frac{1}{2} \mleft[ \lambda \stackrel{\wedge}{,} \lambda \mright]_K \mright)
	- \mleft( \mathrm{ad} \circ \lambda \mright) \wedge \widetilde{\zeta}^\lambda
=
\mathrm{d}^\nabla \zeta
\eas
for all $\lambda \in \Omega^1(N;K)$.
\end{proof}

Let us study this centre-valued form. In fact, $\mathrm{d}^\nabla$ is a differential on centre-valued forms.

\begin{theorems}{Differential on centre-valued forms, \newline \cite[\S 7.2, Definition 7.2.3 and the discussion directly before; page 273]{mackenzieGeneralTheory}}{DifferentialAufZentrumsDinge}
Let $(K, \Xi)$ be a pairing. Then every Lie derivation law $\nabla$ covering $\Xi$ restricts to a flat connection $\nabla^{Z(K)}$ on $Z(K)$.

Moreover, $\Xi$ induces a differential $\mathrm{d}^\Xi: \Omega^\bullet(N; Z(K)) \to \Omega^{\bullet+1}(N; Z(K))$ by choosing $\mathrm{d}^\Xi \coloneqq \mathrm{d}^{\nabla^{Z(K)}} = \mleft.\mathrm{d}^\nabla\mright|_{\Omega^\bullet(N; Z(K))}$ for any Lie derivation law $\nabla$ covering $\Xi$. $\mathrm{d}^\Xi$ is independent of the choice of $\nabla$.

We call this differential \textbf{central representation of $\Xi$}.
\end{theorems}

\begin{proof}
\leavevmode\newline
By Thm.~\ref{thm:GaugeTheoryNeedsLieDerivLawsCoveringACoupling} $\nabla$ satisfies compatibility conditions
\bas
\nabla_Y\mleft( \mleft[ \mu, \nu \mright]_K \mright)
&=
\mleft[ \nabla_Y \mu, \nu \mright]_K
	+ \mleft[ \mu, \nabla_Y \nu \mright]_K, \\
R_\nabla(Y, Z)
&=
\mathrm{ad}(\zeta(Y, Z))
\eas
for all $Y, Z \in \mathfrak{X}(N)$, $\mu, \nu \in \Gamma(K)$ and for some $\zeta \in \Omega^2(N; K)$. Let $\mu \in \Gamma(Z(K))$, then the first compatibility condition implies
\bas
0 &= \mleft[ \nabla_Y \mu, \nu \mright]_K
\eas
for all $Y \in \mathfrak{X}(N)$, $\nu \in \Gamma(K)$ and $\mu \in \Gamma(Z(K))$. That implies that $\nabla_Y \mu \in \Gamma(Z(K))$ such that $\nabla$ is also a connection on $\Gamma(Z(K))$, which we now denote by $\nabla^{Z(K)}$. Restricting the second compatibility condition onto $Z(K)$ then immediately implies
\bas
R_{\nabla^{Z(K)}} &= 0,
\eas
\textit{i.e.}~$\nabla^{Z(K)}$ is flat, and therefore, by the definition of the exterior covariant derivative,
\bas
\mathrm{d}^\Xi &\coloneqq \mleft.\mathrm{d}^\nabla\mright|_{\Omega^\bullet(N; Z(K))} = \mathrm{d}^{\nabla^{Z(K)}}
\eas
is a differential. Now take any other Lie derivation law $\nabla^\prime$ covering $\Xi$.\footnote{Recall the second paragraph of Remark~\ref{remExistenceOfLieDerivationLawsCoveringACoupling}, \textit{i.e.}~there is a Lie derivation Law $\nabla: \mathrm{T}N \to \mathcal{D}_{\mathrm{Der}}(K)$ covering $\Xi$.} By Prop.~\ref{prop:FieldRedefPreservesCoupling}, there is a $\lambda \in \Omega^1(N; K)$ such that
\bas
\nabla^\prime
&=
\nabla - \mathrm{ad} \circ \lambda,
\eas
\textit{i.e.}
\bas
\nabla^\prime_Y \mu
&=
\nabla_Y \mu
\eas
for all $Y \in \mathfrak{X}(N)$ and $\mu \in \Gamma(Z(K))$. Hence, $\mathrm{d}^\Xi$ is independent of the choice of $\nabla$.
\end{proof}


One can now check that $\mathrm{d}^\nabla \zeta$ is closed under $\mathrm{d}^\Xi$. Be aware of that for non-flat Lie derivation laws $\nabla$ covering $\Xi$ this is not a trivial question; due to compatibility condition~\eqref{CondKruemmungmitBLAB}, $\zeta$ is not centre-valued in general such that $\mathrm{d}^\nabla \zeta$ is not the same as $\mathrm{d}^\Xi \zeta$.

\begin{lemmata}{Closedness of $\mathrm{d}^\nabla \zeta$ under the central representation, \newline \cite[\S 7.2, Lemma 7.2.5, $\mathrm{d}^\nabla \zeta$ is denoted by $f$ and $\mathrm{d}^\Xi$ as $d$ there; without written proof; page 274]{mackenzieGeneralTheory}}{DNablaZetaIsClosedUnderDXi}
Let $(K, \Xi)$ be a pairing of $\mathrm{T}N$ and $\nabla$ be a Lie derivation law covering $\Xi$. Also let $\zeta$ be any element of  $\Omega^2(N; K)$ that satisfies compatibility condition~\eqref{CondKruemmungmitBLAB} with respect to $\nabla$.

Then 
\ba
\mathrm{d}^\Xi \mathrm{d}^\nabla \zeta
&=
0
\ea
\textit{i.e.}~$\mathrm{d}^\nabla \zeta \in \Omega^3(N; Z(K))$ is closed under $\mathrm{d}^\Xi$.
\end{lemmata}

\begin{proof}
\leavevmode\newline
We have
\bas
\mleft( \mathrm{d}^\nabla \mright)^2 \zeta
&=
R_\nabla \wedge \zeta
\stackrel{\text{Eq.~\eqref{CondKruemmungmitBLAB}}}{=}
\mleft( \mathrm{ad} \circ \zeta \mright) \wedge \zeta
\stackrel{\text{Eq.~\eqref{wedgeproduktmitadLambdaergibtLieklammer}}}{=}
\mleft[ \zeta \stackrel{\wedge}{,} \zeta \mright]_K,
\eas
but also, using that $\zeta \in \Omega^2(N;K)$,
\bas
\mleft[ \zeta \stackrel{\wedge}{,} \zeta \mright]_K
\stackrel{\text{Eq.~\eqref{VertauschungsregelForKKlammerAufFormen}}}{=}
- \mleft[ \zeta \stackrel{\wedge}{,} \zeta \mright]_K,
\eas
such that $\mleft( \mathrm{d}^\nabla \mright)^2 \zeta = - \mleft( \mathrm{d}^\nabla \mright)^2 \zeta$. Hence, the last statement follows.
\end{proof}

%

We need to know how $\mathrm{d}^\nabla \zeta$ changes by varying $\zeta$.

\begin{lemmata}{Varying $\zeta$ in $\mathrm{d}^\nabla \zeta$, \newline \cite[\S 7.2, Lemma 7.2.6, Mackenzie denotes $\zeta$ by $\Lambda$, $\mathrm{d}^\nabla \zeta$ by $f$ and $\mathrm{d}^\Xi$ by $d$; page 274]{mackenzieGeneralTheory}}{ZetaKannGutGeaendertWerden}
Let $(K, \Xi)$ be a pairing of $\mathrm{T}N$ and $\nabla$ be a Lie derivation law covering $\Xi$. Also let $\zeta$ and $\zeta^\prime$ be two elements of  $\Omega^2(N; K)$ which satisfy compatibility condition~\eqref{CondKruemmungmitBLAB} with respect to $\nabla$.

Then
\ba
\zeta^\prime - \zeta \in \Omega^2(N; Z(K)).
\ea
Especially, $\mathrm{d}^\nabla\zeta^\prime - \mathrm{d}^\nabla\zeta$ is $\mathrm{d}^\Xi$-exact. 
\end{lemmata}

\begin{proof}
\leavevmode\newline
This simply follows by the compatibility condition~\eqref{CondKruemmungmitBLAB}, \textit{i.e.}
\bas
\mleft[ \zeta^\prime(Y, Z) - \zeta(Y, Z), \mu \mright]_K
&=
R_\nabla(Y, Z) \mu - R_\nabla(Y, Z) \mu
= 0
\eas
for all $Y, Z \in \mathfrak{X}(N)$ and $\mu \in \Gamma(K)$. Thence, $\xi \coloneqq \zeta^\prime - \zeta$ is an element of $\Omega^2(N; Z(K))$. By Thm.~\ref{thm:DifferentialAufZentrumsDinge} we get
\bas
\mathrm{d}^\nabla\zeta^\prime - \mathrm{d}^\nabla\zeta
&=
\mathrm{d}^\nabla\underbrace{\mleft(\zeta^\prime - \zeta\mright)}_{\mathclap{\in \Omega^2(N; Z(K))}}
=
\mathrm{d}^\Xi\mleft(\zeta^\prime - \zeta\mright),
\eas
\textit{i.e.}~$\mathrm{d}^\nabla\zeta^\prime - \mathrm{d}^\nabla\zeta$ is exact with respect to $\mathrm{d}^\Xi$ since $\zeta^\prime - \zeta$ has values in $Z(K)$.
\end{proof}

Since $\mathrm{d}^\nabla \zeta$ is invariant under the field redefinition,
%
%
%
%
this finally shows that $\mathrm{d}^\nabla \zeta$ is a useful object to study in the context of the field redefinition. By Lemma~\ref{lem:DNablaZetaIsClosedUnderDXi} this is a closed form, and it is clear that in the flat situation $\zeta$ has values in $Z(K)$ by compatibility condition~\eqref{CondKruemmungmitBLAB}. By Thm.~\ref{thm:DifferentialAufZentrumsDinge} we would get $\mathrm{d}^\nabla \zeta = \mathrm{d}^\Xi \zeta$, \textit{i.e.}~$\mathrm{d}^\nabla \zeta$ would be exact in the flat situation. Hence, it makes sense to study the cohomology class of $\mathrm{d}^\nabla \zeta$ with respect to $\mathrm{d}^\Xi$ if one is interested into whether or not the gauge theory can be transformed into a pre-classical\footnote{Recall Def.~\ref{def:ClassicalGT}.} gauge theory by the field redefinitions.

We denote the space of cohomology classes of $\mathrm{d}^\Xi$-closed elements of $\Omega^\bullet(N; Z(K))$ by
\ba
\mathcal{H}^\bullet\mleft(\mathrm{T}N, \mathrm{d}^\Xi, Z(K)\mright)
\ea
as in \cite[Theorem 7.2.12, replace $A$ with $\mathrm{T}N$ and $\rho^\Xi$ with $\mathrm{d}^\Xi$; page 277]{mackenzieGeneralTheory}, and the classes by $\mleft[ \cdot \mright]_\Xi$. Thus,
\bas
\mleft[ \mathrm{d}^\nabla \zeta \mright]_\Xi &\in \mathcal{H}^3\mleft(\mathrm{T}N, \mathrm{d}^\Xi, Z(K)\mright),
\eas
using that $\mathrm{d}^\nabla \zeta$ is $\mathrm{d}^\Xi$-closed by Lemma~\ref{lem:DNablaZetaIsClosedUnderDXi}.

\begin{theorems}{Cohomology of $\mathrm{d}^\nabla \zeta$ an invariant, \newline \cite[\S 7.2, Theorem 7.2.12, Mackenzie denotes $\mathrm{d}^\Xi$ with $\rho^\Xi$, $\zeta$ with $\Lambda$, $\mathrm{d}^\nabla \zeta$ with $f(\nabla, \Lambda)$, and replace $A$ with $\mathrm{T}N$; page 277]{mackenzieGeneralTheory}}{ObstructionClassIstGeileInvariante}
Let $(K, \Xi)$ be a pairing of $\mathrm{T}N$ and $\nabla$ be a Lie derivation law covering $\Xi$. Also let $\zeta$ be any element of  $\Omega^2(N; K)$ that satisfies compatibility condition~\eqref{CondKruemmungmitBLAB} with respect to $\nabla$.

Then $\mleft[ \mathrm{d}^\nabla \zeta \mright]_\Xi$ only depends on $\Xi$ and not on the particular choice of $\nabla$ and $\zeta$.
\end{theorems}

\begin{proof}
\leavevmode\newline
This follows by Lemma~\ref{lem:ZetaKannGutGeaendertWerden} and Eq.~\eqref{InvarianteFuerFieldRedefImFallLAB}. The former shows that changing $\zeta$ with another element $\zeta^\prime$ of $\Omega^2(N; K)$ satisfying compatibility condition~\eqref{CondKruemmungmitBLAB} results into
\bas
\mathrm{d}^\nabla \zeta^\prime
&=
\mathrm{d}^\nabla \zeta
	+ \underbrace{\mathrm{d}^\nabla \mleft( \zeta^\prime - \zeta \mright)}_{\mathrm{d}^\Xi\text{-exact}}
\in \mleft[ \mathrm{d}^\nabla \zeta \mright]_\Xi,
\eas
\textit{i.e.}~$\mleft[\mathrm{d}^\nabla \zeta^\prime\mright]_\Xi = \mleft[\mathrm{d}^\nabla \zeta \mright]_\Xi$, and the latter shows 
\bas
\mleft[ \mathrm{d}^{\widetilde{\nabla}^\lambda} \widetilde{\zeta}^\lambda \mright]_\Xi
&=
\mleft[ \mathrm{d}^{\nabla} \zeta \mright]_\Xi.
\eas
Thence, by using Prop.~\ref{prop:FieldRedefPreservesCoupling}, \textit{i.e.}~one can reach every other Lie derivation law covering $\Xi$ by using the field redefinition~\ref{fieldredef:FieldRedefForLABs}, one can freely change the Lie derivation law covering $\Xi$ by Eq.~\eqref{InvarianteFuerFieldRedefImFallLAB}, and by Lemma~\ref{lem:ZetaKannGutGeaendertWerden} it doesn't matter which $\zeta$ is used.
\end{proof}

This clearly motivates the following definition of Mackenzie's obstruction class.

\begin{definitions}{The obstruction class of couplings, \newline \cite[\S 7.2, comment after Theorem 7.2.12; page 277]{mackenzieGeneralTheory}}{ObstructionClassOfXi}
Let $(K, \Xi)$ be a pairing of $\mathrm{T}N$, and let $\nabla$ be any Lie derivation law covering $\Xi$. Also let $\zeta$ be any element of  $\Omega^2(N; K)$ that satisfies compatibility condition~\eqref{CondKruemmungmitBLAB} with respect to $\nabla$.

Then we define the \textbf{obstruction class of $\Xi$} by
\ba
\mathrm{Obs}(\Xi)
&\coloneqq
\mleft[ \mathrm{d}^\nabla \zeta \mright]_\Xi.
\ea
\end{definitions}

We immediately get a first result related to CYMH GT.

\begin{corollaries}{First approach of obstruction for CYMH GT on LABs}{FirstApproachOfLABConstruction}
Let $(K, \Xi)$ be a pairing of $\mathrm{T}N$, and let $\nabla$ be a fixed Lie derivation law covering $\Xi$.

Then we have
\bas
\exists \text{ a field redefinition as in~\ref{fieldredef:FieldRedefForLABs}}: ~ \widetilde{\nabla}^\lambda \text{ is flat}
\quad&\Rightarrow\quad
\mathrm{Obs}(\Xi) = 0\in\mathcal{H}^3\mleft(\mathrm{T}N, \mathrm{d}^\Xi, Z(K)\mright).
\eas
Or, equivalently, if there is a flat Lie derivation law covering $\Xi$ then $\mathrm{Obs}(\Xi) = 0$.
\end{corollaries}
%

\begin{proof}
\leavevmode\newline
Let $\zeta$ be any element of $\Omega^2(N; K)$ that satisfies compatibility condition~\eqref{CondKruemmungmitBLAB} with respect to $\nabla$. When there is a field redefinition such that $\widetilde{\nabla}^\lambda$ is flat then we can conclude that $\widetilde{\zeta}^\lambda$ has only values in $Z(K)$ by compatibility condition~\eqref{CondKruemmungmitBLAB}. But then we arrive at
\bas
\mathrm{Obs}(\Xi)
&=
\mleft[ \mathrm{d}^\nabla \zeta \mright]_\Xi
\stackrel{\text{Eq.~\eqref{InvarianteFuerFieldRedefImFallLAB}}}{=}
\mleft[ \mathrm{d}^{\widetilde{\nabla}^\lambda} \widetilde{\zeta}^\lambda \mright]_\Xi
\stackrel{\text{Thm.~\ref{thm:DifferentialAufZentrumsDinge}}}{=}
\mleft[ \mathrm{d}^{\Xi} \widetilde{\zeta}^\lambda \mright]_\Xi
=
0.
\eas
The last statement simply follows by Prop.~\ref{prop:FieldRedefPreservesCoupling}.
\end{proof}

\subsection{Mackenzie's theory about extensions of tangent bundles}

We now want to study when the obstruction is zero and when it implies the existence of a flat Lie derivation law covering $\Xi$. To understand this, we need to understand why Mackenzie studied this obstruction class and got similar formulas. Mackenzie was interested into whether or not a Lie algebroid can be extended by an LAB; we are going to state Mackenzie's statements in the special situation of having $\mathrm{T}N$ as the Lie algebroid. But the arguments and calculations do not really differ; in the context of gauge theory we just need to study $\mathrm{T}N$. Many proofs are very straightforward and/or extremely long, so, we refer to the references for the proofs, we will not repeat them.

\begin{definitions}{Extension of tangent bundles by LABs and transversals, \newline \cite[\S 7.1, Definition 7.1.11, page 266, and Definition 7.3.1, page 277]{mackenzieGeneralTheory}}{ExtensionOfTNByLABs}
Let $K \to N$ be an LAB. Then an \textbf{extension of $\mathrm{T}N$ by $K$} is a short exact sequence of Lie algebroids over $N$
\be\label{defShortExactSeqExtensionOfTNByK}
	\begin{tikzcd}
		K \arrow[hook]{r}{\iota} & E \arrow[two heads]{r}{\pi} & \mathrm{T}N.
	\end{tikzcd}
\ee
A transversal of~\eqref{defShortExactSeqExtensionOfTNByK} is a vector bundle morphism $\chi: \mathrm{T}N \to E$ such that $\pi \circ \chi = \mathds{1}_{\mathrm{T}N}$.
\end{definitions}

\begin{remark}
\leavevmode\newline
We will, as usual, denote the Lie bracket of $E$ by $\mleft[ \cdot, \cdot \mright]_E$, and $\pi$ is its anchor due to the fact that $\pi$ is anchor-preserving.
\end{remark}

To a given transversal we are able to define a Lie derivation law covering some Lie algebroid morphism $\Xi: \mathrm{T}N \to \mathrm{Out}(\mathcal{D}_{\mathrm{Der}}(K))$.

\begin{propositions}{Lie derivation law of a transversal, \newline \cite[\S 7.3, Proposition 7.3.2 and Lemma 7.3.3, replace $A$ with $\mathrm{T}N$ and $A^\prime$ with $E$; page 278]{mackenzieGeneralTheory}}{TransversalAndItsLieDerivationLaw}
Let
\begin{center}
	\begin{tikzcd}
		K \arrow[hook]{r}{\iota} & E \arrow[two heads]{r}{\pi} & \mathrm{T}N
	\end{tikzcd}
\end{center}
be an extension of $\mathrm{T}N$ by an LAB $K \to N$, and let $\chi$ be any transversal. Then a connection $\nabla^\chi$ on $K$, given by
\ba\label{DefTransversalConnection}
\iota\mleft( \nabla^\chi_Y \mu \mright)
&=
\mleft[ \chi(Y), \iota(\mu) \mright]_E
\ea
for all $Y \in \mathfrak{X}(N)$ and $\mu \in \Gamma(K)$, describes a Lie derivation law covering some Lie algebroid morphism $\Xi: \mathrm{T}N \to \mathrm{Out}(\mathcal{D}_{\mathrm{Der}}(K))$.
\end{propositions}

Furthermore, the Lie algebroid morphism covered by $\nabla^\chi$ is the same for all transversals $\chi$.

\begin{corollaries}{All transversals results into the same covered pairing, \newline \cite[\S 7.3, comment after Lemma 7.3.3, replace $A$ with $\mathrm{T}N$ and $A^\prime$ with $E$; page 278]{mackenzieGeneralTheory}}{TransversalsCoverTheSameCoupling}
Let
\begin{center}
	\begin{tikzcd}
		K \arrow[hook]{r}{\iota} & E \arrow[two heads]{r}{\pi} & \mathrm{T}N
	\end{tikzcd}
\end{center}
be an extension of $\mathrm{T}N$ by an LAB $K \to N$, and let $\chi$ and $\chi^\prime$ be two transversals.

Then
\bas
\sharp \circ \nabla^\chi
&=
\sharp \circ \nabla^{\chi^\prime}.
\eas
\end{corollaries}
%

This immediately leads to the following definition.

\begin{definitions}{Pairing induced by an extension, \newline \cite[\S7.3, Definition 7.3.4, replace $A$ with $\mathrm{T}N$ and $A^\prime$ with $E$; page 278]{mackenzieGeneralTheory}}{CouplingsOfExtensions}
Let
\begin{center}
	\begin{tikzcd}
		K \arrow[hook]{r}{\iota} & E \arrow[two heads]{r}{\pi} & \mathrm{T}N
	\end{tikzcd}
\end{center}
be an extension of $\mathrm{T}N$ by an LAB $K \to N$, and let $\chi$ be any transversal.

Then the pairing $\Xi_{\mathrm{ext}} \coloneqq \sharp \circ \nabla^\chi: \mathrm{T}N \to \mathrm{Out}\mleft( \mathcal{D}_{\mathrm{Der}}(K) \mright)$ is the \textbf{pairing of $\mathrm{T}N$ with $K$ induced by the extension}.
\end{definitions}

Finally we can state what Mackenzie has shown about the obstruction class.

\begin{theorems}{Obstruction of an extension, \newline \cite[\S 7.3, Proposition 7.3.6, page 279, Corollary 7.3.9 and the comment afterwards, page 281; replace $A$ with $\mathrm{T}N$ and $A^\prime$ with $E$]{mackenzieGeneralTheory}}{ObstructionOfExtensions}
Let $(K, \Xi)$ be a pairing of $\mathrm{T}N$.

Then there is an extension
\begin{center}
	\begin{tikzcd}
		K \arrow[hook]{r}{\iota} & E \arrow[two heads]{r}{\pi} & \mathrm{T}N
	\end{tikzcd}
\end{center}
of $\mathrm{T}N$ by $K$ such that $\Xi_{\mathrm{ext}} = \Xi$ if and only if $\mathrm{Obs}(\Xi) = 0 \in \mathcal{H}^3\mleft(\mathrm{T}N, \mathrm{d}^\Xi, Z(K)\mright)$. Moreover, given such an extension, then for all Lie derivation laws $\nabla$ covering $\Xi$ there is a transversal $\chi$ such that
\bas
\nabla
&=
\nabla^\chi.
\eas
\end{theorems}

By Cor.~\ref{cor:FirstApproachOfLABConstruction} we see that the question about whether there is a field redefinition in sense of~\ref{fieldredef:FieldRedefForLABs} to arrive at a pre-classical gauge theory, \textit{i.e.}~when $\nabla$ is flat, is related to the existence of an extension of $\mathrm{T}N$ by $K$.

When we are just interested into local behaviours then we might assume that $N$ is contractible.

\begin{theorems}{Extensions over contractible manifolds, \newline \cite[\S 8.2, Theorem 8.2.1, replace $A$ with $E$, $L$ with $K$ and $TM$ with $\mathrm{T}N$; page 314]{mackenzieGeneralTheory}}{ExtensionWennNContrahierbar}
Let
\begin{center}
	\begin{tikzcd}
		K \arrow[hook]{r}{\iota} & E \arrow[two heads]{r}{\pi} & \mathrm{T}N
	\end{tikzcd}
\end{center}
be an extension of $\mathrm{T}N$ by an LAB $K$ over a contractible manifold $N$. Then there is a flat Lie derivation law covering $\Xi_{\mathrm{Ext}}$.\footnote{Mackenzie stated that $E$ admits a flat connection, with that they actually mean that it is a flat Lie derivation law covering $\Xi_{\mathrm{Ext}}$.}
\end{theorems}

\subsection{Results}

In total we derive therefore the following two statements, the first can be seen as a generalization of Cor.~\ref{cor:CorLocalerFlacherZusammenhangFuerIrgendeineKopplung}.

\begin{theorems}{Local existence of pre-classical gauge theory}{LokalLeiderImmerPreklassisch}
Let $(K, \Xi)$ be a pairing of $\mathrm{T}N$ over a contractible manifold $N$, and let $\nabla$ be a fixed Lie derivation law covering $\Xi$.

Then we have a field redefinition in sense of~\ref{fieldredef:FieldRedefForLABs}, \textit{i.e.}~there is a $\lambda\in\Omega^1(N;K)$ such that $\widetilde{\nabla}^\lambda$ is flat.
\end{theorems}

\begin{proof}
\leavevmode\newline
We only need to show that $\mathrm{Obs}(\Xi) = \mleft[ \mathrm{d}^\nabla \zeta \mright]_\Xi = 0$, where $\zeta \in \Omega^2(N; K)$ such that compatibility condition~\eqref{CondKruemmungmitBLAB} is satisfied. As given in Thm.~\ref{thm:DifferentialAufZentrumsDinge} the central representation $\mathrm{d}^\Xi$ of $\Xi$ is basically $\mathrm{d}^{\nabla^{Z(K)}}$ where $\nabla^{Z(K)}$ is $\nabla$ restricted on the subbundle $Z(K)$, and we have shown that $\nabla^{Z(K)}$ is flat by compatibility condition~\eqref{CondKruemmungmitBLAB}. Due to the fact that $N$ is contractible, we have a global parallel frame $\mleft( e_a \mright)_a$ for $Z(K)$ with respect to $\nabla^{Z(K)}$.

By Prop.~\ref{thm:BianchiIdentityForZeta} we have $\mathrm{d}^\nabla \zeta \in \Omega^3(N; Z(K))$, thence, we can write $\mathrm{d}^\nabla \zeta = \omega^a \otimes e_a$ with $\omega^a \in \Omega^3(N)$. We arrive at
\bas
\mathrm{d}^\Xi \mathrm{d}^\nabla \zeta
&=
\mathrm{d}\omega^a \otimes e_a,
\eas
where $\mathrm{d}$ is the standard deRham differential. So, the differential breaks down to the standard differential in each component, especially closedness and exactness mean to be closed and exact in each component with respect to $\mleft( e_a \mright)_a$, respectively. By Lemma~\ref{lem:DNablaZetaIsClosedUnderDXi} we have $\mathrm{d}^\Xi \mathrm{d}^\nabla \zeta=0$, thus, $\mathrm{d}\omega^a = 0$. Again due to that $N$ is contractible, we can conclude that closedness implies exactness by the Poincaré lemma. Thence, $\mathrm{Obs}(\Xi) = 0$.

By Thm.~\ref{thm:ObstructionOfExtensions} we have an extension
\begin{center}
	\begin{tikzcd}
		K \arrow[hook]{r}{\iota} & E \arrow[two heads]{r}{\pi} & \mathrm{T}N.
	\end{tikzcd}
\end{center}
such that $\Xi_{\mathrm{ext}} = \Xi$, and, hence, a flat Lie derivation law covering $\Xi$ by Thm.~\ref{thm:ExtensionWennNContrahierbar}. By Prop.~\ref{prop:FieldRedefPreservesCoupling} the existence of the field redefinition to a flat derivation law covering $\Xi$ follows.
\end{proof}

\begin{theorems}{Possible new and curved gauge theories on LABs}{NeueLABGTs}
Let $(K, \Xi)$ be a pairing of $\mathrm{T}N$ with $\mathrm{Obs}(\Xi) \neq 0$ and such that the fibre Lie algebra $\mathfrak{g}$ admits an $\mathrm{ad}$-invariant scalar product.

Then we can construct a CYMH GT for which there is no field redefinition with what it would become pre-classical.
\end{theorems}

\begin{proof}
\leavevmode\newline
Take any Lie derivation law $\nabla$ covering $\Xi$ (recall the second paragraph of Remark~\ref{remExistenceOfLieDerivationLawsCoveringACoupling} about the existence of $\nabla$ for a given $\Xi$). By Thm.~\ref{thm:GaugeTheoryNeedsLieDerivLawsCoveringACoupling} this connection satisfies compatibility conditions~\eqref{CondSGleichNullLAB} and~\eqref{CondKruemmungmitBLAB}. Together with the existence of an $\mathrm{ad}$-invariant scalar product we have everything what we need to construct a CYMH GT in sense of~\ref{sit:CYMHGTForLABsToDoList}.

Due to $\mathrm{Obs}(\Xi) \neq 0$ and Cor.~\ref{cor:FirstApproachOfLABConstruction} the statement follows.
\end{proof}

Hence, we have shown that $\mathrm{Obs}(\Xi)$ is not just an obstruction for extensions of $\mathrm{T}N$, it also leads to an obstruction for the question about whether or not a CYMH GT can be transformed to a pre-classical gauge theory by a field redefinition. However, Mackenzie also has shown that there are examples with zero obstruction class but without a flat Lie derivation law covering the pairing. Thus, there is in general only for contractible $N$ an equivalence of $\mathrm{Obs}(\Xi) = 0$ and the existence of flat Lie derivation laws covering a pairing.

\begin{examples}{The isotropy of a Hopf bundle, \cite[Example 7.3.20; page 287]{mackenzieGeneralTheory}}{HopfBuendelEventuellSuperFragezeichen}
Let $P$ be the Hopf bundle
\begin{center}
	\begin{tikzcd}
		\mathrm{SU}(2) \arrow{r}	& \mathds{S}^7 \arrow{d} \\
			& \mathds{S}^4
	\end{tikzcd}
\end{center}
Then for the adjoint bundle
\bas
K
&\coloneqq
P \times_{\mathrm{SU}(2)} \mathrm{su}(2)
\coloneqq 
\mleft( \mathds{S}^7 \times \mathfrak{su}(2) \mright) \Big/ \mathrm{SU}(2)
\eas
we have the Atiyah sequence
\begin{center}
	\begin{tikzcd}
		K \arrow[hook]{r}{\iota} & \mathrm{T}P \Big/ \mathrm{SU}(2) \arrow[two heads]{r}{\pi} & \mathrm{T}\mathds{S}^4.
	\end{tikzcd}
\end{center}
of $\mathrm{T}\mathds{S}^4$ by $K$. We can view this sequence as an extension.

Then $\mathrm{Obs}(\Xi_{\mathrm{Ext}}) = 0$, but there is no flat derivation law, especially no flat derivation law covering $\Xi_{\mathrm{Ext}}$.
\end{examples}

\begin{remark}
\leavevmode\newline
The fibre of $K$ is given by $\mathrm{su}(2)$, and, thence, the existence of an $\mathrm{ad}$-invariant scalar product is given. Therefore this gives an example of a CYMH GT as in~\ref{sit:CYMHGTForLABsToDoList} by taking any fibre metric $\kappa$ on $K$ which restricts to an $\mathrm{ad}$-invariant scalar product on each fibre, and taking any Lie derivation law $\nabla$ covering $\Xi_{\mathrm{Ext}}$ such that the existence of a $\zeta \in \Omega^2(N;K)$ as in compatibility condition~\eqref{CondKruemmungmitBLAB} is given. By Prop.~\ref{prop:FieldRedefPreservesCoupling} this example shows that there is no field redefinition as in~\ref{fieldredef:FieldRedefForLABs} such that this gauge theory would become pre-classical.

Observe that a trivial semisimple LAB would not work: Fix any global frame $\mleft( e_a \mright)_a$ of the trivial LAB, then we would have $\nabla e_a = \mleft[ \lambda , e_a \mright]_K$ for a $\lambda \in \Omega^1(N; K)$ because all bracket derivations are inner derivations for semisimple Lie algebras; for this, simply view the connection 1-forms $\omega_a^b$, given by $\nabla e_a = \omega_a^b \otimes e_b$, as matrices acting on constant (w.r.t. $\mleft( e_a \mright)_a$) sections. Then $\widetilde{\nabla}^\lambda$ would be flat, and its parallel frame is \textit{e.g.}~given by $\mleft( e_a \mright)_a$. This argument just depends on the triviality of the LAB, regardless whether the base is contractible or not. The obstruction class is of course always trivial for semisimple LABs because their centre is zero.
\end{remark}

\section{\texorpdfstring{Existence of non-vanishing $\zeta$ stable under the field redefinition}{Existence of non-vanishing 2-forms with values in the LAB contributing to the field strength, stable under the field redefinition}} \label{NonclassicalStuff}

When one is interested into perturbation theory, especially just in a local theory, then Thm.~\ref{thm:LokalLeiderImmerPreklassisch} seems to show that locally one can not hope for new gauge theories, especially ones related to non-flat $\nabla$. However, we still have the two-form $\zeta$. As argued in Rem.~\ref{rem:MotivationForTheNewFieldStrength} in the situation of abelian $K$ there is the existence of gauge theories which are never classical, not even locally, by choosing $\mathrm{d}^\nabla \zeta \neq 0$. We can transform them locally to pre-classical ones by Thm.~\ref{thm:LokalLeiderImmerPreklassisch} but surely not always to classical ones.

\begin{theorems}{Existence of LABs giving rise to non-classical gauge theories}{AbelschIstGeileNeueTheorie}
Let $K \to N$ be an LAB , $\nabla$ a connection satisfying compatibility conditions~\eqref{CondSGleichNullLAB} and~\eqref{CondKruemmungmitBLAB} with respect to a given $\zeta \in \Omega^2(N; K)$ such that $\mathrm{d}^\nabla \zeta \neq 0$.

Then there is no $\lambda\in\Omega^1(N;K)$ as in~\ref{fieldredef:FieldRedefForLABs} such that $\widetilde{\zeta}^\lambda = 0$.
\end{theorems}

\begin{proof}
\leavevmode\newline
We have a 2-form $\zeta \in \Omega^2(N; K)$ such that
\bas
\mathrm{d}^\nabla \zeta &\neq 0.
\eas
By Eq.~\eqref{InvarianteFuerFieldRedefImFallLAB} we have $\mathrm{d}^{\widetilde{\nabla}^\lambda} \widetilde{\zeta}^\lambda= \mathrm{d}^\nabla \zeta$ for all $\lambda \in \Omega^1(N;K)$. When there would be a field redefinition leading to a classical gauge theory, then $\widetilde{\zeta}^\lambda = 0$ but then also $\mathrm{d}^{\widetilde{\nabla}^\lambda} \widetilde{\zeta}^\lambda = 0$. Thence, by $\mathrm{d}^\nabla \zeta \neq 0$ the statement follows.
\end{proof}

Starting with a standard Yang-Mills gauge theory with an additional free physical field $X$ with a Lagrangian similar to the Higgs field, we have a canonical construction when the centre of the Lie algebra is non-trivial.

\begin{corollaries}{Canonical construction of non-classical gauge theories}{CanonicalConstructionOfGaugeTheories}
Let $\mathfrak{g}$ be a Lie algebra with non-zero centre and admitting an $\mathrm{ad}$-invariant scalar product. Also let $(N, g)$ be any Riemannian manifold with at least three dimensions, and $K = N \times \mathfrak{g}$ be a trivial LAB over $N$, equipped with the canonical flat connection $\nabla$ and a metric $\kappa$ which restricts to an $\mathrm{ad}$-invariant scalar product on each fibre.

Then there is a $\zeta \in \Omega^2(N; Z(K))$ in sense of~\ref{sit:CYMHGTForLABsToDoList}, with $\mathrm{d}^\nabla \zeta \neq 0$, such that this set-up describes a non-classical YMH GT with respect to an arbitrary spacetime $M$. Additionally, there is no $\lambda\in\Omega^1(N;K)$ as in~\ref{fieldredef:FieldRedefForLABs} such that $\widetilde{\zeta}^\lambda = 0$.
\end{corollaries}

\begin{proof}
\leavevmode\newline
By the assumptions we have everything we need to formulate a YMH GT for a given spacetime $M$, following~\ref{sit:CYMHGTForLABsToDoList}; as in the third paragraph of Remark~\ref{WePrueftManDenIsomorphismusZuStandardTheorie}, compatibility condition~\eqref{CondSGleichNullLAB} follows by testing this condition with respect to a constant frame. For compatibility condition~\eqref{CondKruemmungmitBLAB} just take any element of $\Omega^2(N; Z(K))$, denoted as $\zeta$, then this condition is trivially satisfied because $\nabla$ is flat and $\zeta$ only has values in the centre of $K$.

Since $N$ is three-dimensional and $Z(K)$ is non-zero, we can then conclude the existence of $\mathrm{d}^\nabla \zeta \neq 0$. For this recall that $\mathrm{d}^\nabla \zeta$ is still a centre-valued form by Eq.~\eqref{dnablazetaIsCentrevalued} and that $\mathrm{d}^\nabla$ is then just the differential $\mathrm{d}^\Xi$ for $\Xi \coloneqq \sharp \circ \nabla$ as in Thm.~\ref{thm:DifferentialAufZentrumsDinge}. Therefore we only need to take any non-$\mathrm{d}^\Xi$-closed centre-valued form $\zeta$, of which there are plenty. The non-existence of a $\lambda$ with $\widetilde{\zeta}^\lambda = 0$ then follows by Thm.~\ref{thm:AbelschIstGeileNeueTheorie}.
\end{proof}

\section{The Bianchi identity of the new field strength} \label{BianchiStuff}

We conclude this paper with an interpretation of $\mathrm{d}^\nabla \zeta$, and for this we need to calculate the Bianchi identity of the field strength. Hence, we need to understand how $X^*\nabla$ behaves.

\begin{propositions}{Pull-Back of a Lie derivation law covering a pairing}{PullbackvonUnseremGaugeNabla}
Let $K \to N$ be an LAB, equipped with a connection $\nabla$ satisfying compatibility condition~\eqref{CondSGleichNullLAB}; also let $M$ be another smooth manifold and $X: M \to N$ a smooth map. Then $X^*\nabla$ also satisfies compatibility condition~\eqref{CondSGleichNullLAB} with respect to $X^*K$.
\newline

When $\nabla$ satisfies compatibility condition~\eqref{CondKruemmungmitBLAB} with respect to a $\zeta \in \Omega^2(N; K)$, not necessarily assuming~\eqref{CondSGleichNullLAB}, then this extends to $X^*K$, too, \textit{i.e.}
\ba\label{EqCompCondFuerPullbackCurvature}
R_{X^*\nabla} = \mathrm{ad}^* \circ X^!\zeta,
\ea
viewing the curvature as an element of $\Omega^2(M; X^*K)$.
\end{propositions}

\begin{remark}
\leavevmode\newline
By Thm.~\ref{thm:GaugeTheoryNeedsLieDerivLawsCoveringACoupling}, we get that the pull-back of a Lie derivation law of $K$ covering the Lie algebroid morphism $\sharp \circ \nabla$ is a Lie derivation law of $X^*K$ covering the Lie algebroid morphism $\sharp \circ X^*\nabla$.
\end{remark}

\begin{proof}
\leavevmode\newline
\indent$\bullet$ We can show
\bas
X^*\nabla \underbrace{\mleft( \mleft[ X^*\mu, X^*\nu \mright]_{X^*K} \mright)}
_{=~ X^*\mleft( \mleft[ \mu, \nu \mright]_K \mright)}
~~~&\stackrel{\mathclap{\text{Eq.~\eqref{eqShortNotationForPullbackConnections}}}}{=}~~~
X^!\mleft( \nabla\mleft( \mleft[ \mu, \nu \mright]_K \mright) \mright)\\
&\stackrel{\mathclap{\text{Eq.~\eqref{CondSGleichNullLAB}}}}{=}~~
X^!\mleft(  \mleft[ \nabla\mu, \nu \mright]_K + \mleft[ \mu, \nabla\nu \mright]_K \mright)\\
&\stackrel{\mathclap{\text{Eq.~\eqref{eqPullbackofLiebracketStuff}}}}{=}~~~
\mleft[ X^!(\nabla\mu), X^*\nu \mright]_{X^*K} + \mleft[ X^*\mu, X^!(\nabla\nu) \mright]_{X^*K} \\
&\stackrel{\mathclap{\text{Eq.~\eqref{eqShortNotationForPullbackConnections}}}}{=}~~~
\mleft[ (X^*\nabla)(X^*\mu), X^*\nu \mright]_{X^*K} + \mleft[ X^*\mu, (X^*\nabla)(X^*\nu) \mright]_{X^*K}
\eas
for all $\mu, \nu \in \Gamma(K)$. Since pull-backs of $\Gamma(K)$ generate $\Gamma(X^*K)$ and since~\eqref{CondSGleichNullLAB} is a tensorial equation, we can derive that $X^*\nabla$ also satisfies compatibility condition~\eqref{CondSGleichNullLAB} with respect to the LAB $X^*K$. 

$\bullet$ Now let $\nabla$ satisfy compatibility condition~\eqref{CondKruemmungmitBLAB}, and recall that in general curvatures satisfy
\bas
R_\nabla \nu &= \mleft( \mathrm{d}^\nabla \mright)^2 \nu \in \Omega^2(N; K)
\eas
for all $\nu \in \Gamma(K)$ (see also \cite[\S 5, third part of Exercise 5.15.12; page 316]{hamilton}). Then apply Eq.~\eqref{EqGeilePullBackCommuteFormel} to get
\bas
R_{X^*\nabla} (X^*\nu)
&=
\mleft( \mathrm{d}^{X^*\nabla} \mright)^2 (X^*\nu)
=
X^!\mleft( \mleft(\mathrm{d}^\nabla\mright)^2 \nu \mright)
\stackrel{\text{Eq.~\eqref{CondKruemmungmitBLAB}}}{=}
X^!\mleft( \mleft[ \zeta, \nu \mright]_K \mright)
\stackrel{\text{Eq.~\eqref{eqPullbackofLiebracketStuff}}}{=}
\mleft[ X^!\zeta, X^*\nu \mright]_{X^*K},
\eas
such that $R_{X^*\nabla} = \mathrm{ad}^* \circ X^!\zeta$ follows, by using again that pull-backs of $\Gamma(K)$ generate $\Gamma(X^*K)$.
\end{proof}

Using this we calculate the Bianchi identity for the field strength $G$.

\begin{theorems}{Bianchi identity of the field strength}{BianchiIdentityOfFieldStrength}
Let $M$ and $N$ be smooth manifolds, $K \to N$ an LAB, $X: M \to N$ a smooth map, and $\nabla$ a connection satisfying compatibility conditions~\eqref{CondSGleichNullLAB} and~\eqref{CondKruemmungmitBLAB} with respect to a given $\zeta \in \Omega^2(N; K)$.

Then
\ba
\mathrm{d}^{X^*\nabla}G + \mleft[ A \stackrel{\wedge}{,} G \mright]_{X^*K}
&=
X^! \mleft( \mathrm{d}^\nabla \zeta \mright),
\ea
where
\bas
G
&=
\mathrm{d}^{X^*\nabla}A
	+ \frac{1}{2} \mleft[ A \stackrel{\wedge}{,} A \mright]_{X^*K}
	+ X^!\zeta
\eas
is the field strength of a $A \in \Omega^1(M; X^*K)$.
\end{theorems}

\begin{remark}
\leavevmode\newline
This clearly generalizes the standard Bianchi identity for field strengths, as \textit{e.g.}~given in \cite[\S 5, Theorem 5.14.2; page 311]{hamilton}: Take a trivial LAB $K$ equipped with its canonical flat connection and $\zeta \equiv 0$. Then we arrive at the typical Bianchi identity. In general, we get $\mathrm{d}^{X^*\nabla}G + \mleft[ A \stackrel{\wedge}{,} G \mright]_{X^*K}=0$ when $\mathrm{d}^\nabla \zeta = 0$, which resembles strongly the standard Bianchi identity, but covariantized. Hence, we say that $G$ satisfies the \textbf{Bianchi identity} if and only if $\mathrm{d}^{X^*\nabla}G + \mleft[ A \stackrel{\wedge}{,} G \mright]_{X^*K}=0$.
\end{remark}

\begin{proof}
\leavevmode\newline
The calculation is similarly to the standard calculation of the standard formulation of the Bianchi identity as in \cite[\S 5, Theorem 5.14.2; page 311]{hamilton}, making use of compatibility condition~\eqref{CondSGleichNullLAB} needed for Eq.~\eqref{eqDerivationOfDifferentialOnBracketonK}. We have, viewing the curvature $R_{X^*\nabla}$ as an element of $\Omega^2(M; \mathrm{End}(X^*K))$,
\bas
&\mleft( \mathrm{d}^{X^*\nabla} \mright)^2 A
=
R_{X^*\nabla} \wedge A
\stackrel{\text{Eq.~\eqref{EqCompCondFuerPullbackCurvature}}}{=}
\mleft( \mathrm{ad}^* \circ X^!\zeta \mright) \wedge A
\stackrel{\text{Eq.~\eqref{wedgeproduktmitadLambdaergibtLieklammer}}}{=}
\mleft[ X^!\zeta \stackrel{\wedge}{,} A \mright]_{X^*K}
\stackrel{\text{Eq.~\eqref{VertauschungsregelForKKlammerAufFormen}}}{=}
- \mleft[ A \stackrel{\wedge}{,} X^!\zeta \mright]_{X^*K}, \\
&\mathrm{d}^{X^*\nabla}\mleft( \mleft[ A \stackrel{\wedge}{,} A \mright]_{X^*K} \mright)
\stackrel{\text{Eq.~\eqref{eqDerivationOfDifferentialOnBracketonK}}}{=}
\mleft[ \mathrm{d}^{X^*\nabla} A \stackrel{\wedge}{,} A \mright]_{X^*K}
	- \mleft[ A \stackrel{\wedge}{,} \mathrm{d}^{X^*\nabla} A \mright]_{X^*K}
\stackrel{\text{Eq.~\eqref{VertauschungsregelForKKlammerAufFormen}}}{=}
- 2 ~ \mleft[ A \stackrel{\wedge}{,} \mathrm{d}^{X^*\nabla} A \mright]_{X^*K}, \\
&\mleft[ A \stackrel{\wedge}{,} \mleft[ A \stackrel{\wedge}{,} A \mright]_{X^*K} \mright]_{X^*K}
\stackrel{\text{Eq.~\eqref{JacobiIdentityForFormBracket}}}{=}
0, \\
&\mathrm{d}^{X^*\nabla} \mleft( X^!\zeta \mright)
\stackrel{\text{Eq.~\eqref{EqGeilePullBackCommuteFormel}}}{=}
X^! \mleft( \mathrm{d}^\nabla \zeta \mright),
\eas
and, using all of these, we arrive at
\bas
\mathrm{d}^{X^*\nabla}G + \mleft[ A \stackrel{\wedge}{,} G \mright]_{X^*K}
~~~&\stackrel{\mathclap{\text{Def.~\eqref{defNewFieldStrengthG}}}}{=}~~~
X^! \mleft( \mathrm{d}^\nabla \zeta \mright).
\eas
\end{proof}

Thence, $\mathrm{d}^\nabla \zeta$ measures the failure of the Bianchi identity of the field strength $G$. For example, applying Cor.~\ref{cor:CanonicalConstructionOfGaugeTheories} to the Yang-Mills gauge theory of electromagnetism, \textit{i.e.}~the Lie algebra is given by $\mathfrak{g} = \mathrm{u}(1)$, would result into a gauge theory where there is no (vector) potential of the field strength as usual, so, $G$ could not be written as $\mathrm{d}^\nabla \widehat{A}$ for some $\widehat{A} \in \Omega^1(N;X^*K)$.\footnote{Recall that $\mathrm{d}^\nabla$ is a differential since $\nabla$ is flat in that situation.}

\section{Conclusion}

We have restated a covariantized theory of gauge theory, allowing non-flat vector bundle connections $\nabla$ on the Lie algebra with an additional 2-form $\zeta$ in the field strength, originally introduced by Alexei Kotov and Thomas Strobl, but here just using Lie algebra bundles which resembles Yang-Mills-Higgs gauge theories but without minimal coupling of the gauge bosons to the Higgs field, \textit{i.e.}~the context is here given by massless gauge bosons. Additionally, there is a field redefinition which keeps the Lagrangian invariant, but which might lead to a standard formulation of gauge theory. With this redefinition it is also possible to motivate $\zeta$, then it is not just an auxiliary map allowing non-flat connections by compatibility condition \eqref{CondKruemmungmitBLAB}.

Using the compatibility conditions, we were able to see that the studied connection $\nabla$ is a Lie derivation law covering a Lie algebroid morphism $\Xi: \mathrm{T}N \to \mathrm{Out}(\mathcal{D}_{\mathrm{Der}}(K))$, the latter then given by $\Xi = \sharp \circ \nabla$.

We have seen that the field redefinitions and the question about, whether or not we can transform to a pre-classical gauge theory, where $\nabla$ is flat, has a strong relationship to Mackenzie's study about extending tangent bundles with LABs over the same base in sense of Lie algebroids. Using Mackenzie's results, we were able to use an obstruction class $\mathrm{Obs}(\Xi)$ and to argue that locally we can always transform to a pre-classical gauge theory, while globally it is a different question: Having a non-trivial obstruction class leads quickly to a non-pre-classical gauge theory, while even a trivial obstruction class can still imply a non-pre-classical theory.

The obstruction class is also strongly related to $\mathrm{d}^\nabla \zeta$, an invariant of the field redefinition. Studying this object leads to the quick result that $\mathrm{d}^\nabla \zeta \neq 0$ already implies that there is no field redefinition leading to a classical theory because then $\zeta$ cannot vanish after any field redefinition. This condition implies the failure of the (covariantized) Bianchi identity of the field strength. Moreover, there is a canonical construction of such a theory with $\mathrm{d}^\nabla \zeta \neq 0$ when starting with a classical theory.

But all of this really needs the additional free physical field $X$ without minimal coupling. Of course, (C)YMH GT is also formulated with minimal coupling, making use of general Lie algebroids, especially with non-zero anchor. Internally, the concepts presented here are mostly already generalized and will be presented in another paper.

\textbf{Acknowledgements:} I want to thank Mark John David Hamilton, Anna Dall'Acqua, Alessandra Frabetti, Anton Alekseev and Maxim Efremov for their great help and support in making this paper and my Ph.D. 

This paper is part of my Ph.D.~at two universities (that type of Ph.D.~is called cotutelle), supervised by Anton Alekseev (Universit\'{e} de Gen\`eve) and Thomas Strobl (Universit\'{e} Claude Bernard Lyon 1). In~\cite[the example at the very end before the conclusion; the $B$ there is the $\zeta$ presented here]{CurvedYMH} a transformation got presented which can make $\zeta$ vanish and $\nabla$ flat. This is a special example of the field redefinition which Thomas Strobl got after a private dialogue with Edward Witten. Thomas Strobl suggested Eq.~\eqref{TrafoFormelFuerDieEichbosonenA}, and the aim of my Ph.D. project was to derive all other formulas needed for the field redefinition (in the situation of general Lie algebroids, not just LABs) and to study them. I will present other results in a separate paper, especially regarding the situation regarding general Lie algebroids.

This publication was produced within the scope of the NCCR SwissMAP which was funded by the Swiss National Science Foundation. I would like to thank the Swiss National Science Foundation for their financial support.

This work was supported by the LABEX MILYON (ANR-10-LABX-0070) of Universit\'{e} de Lyon, within the program "Investissements d'Avenir" (ANR-11-IDEX- 0007) operated by the French National Research Agency (ANR).


\newpage



\appendix
\setcounter{equation}{0}
\renewcommand{\theequation}{\Alph{section}.\arabic{equation}} 
\section{Identities for the calculus given in (C)YMH GT}\label{CalculusIdentitiesNeeded}

\begin{propositions}{Several useful identities}{SeveralIdentitiesFortheCalculusWithPullbackandBlah}
Let $M$ and $N$ be two smooth manifolds, $K \to N$ a vector bundle, $X: M \to N$ a smooth map, $\nabla$ a connection on $K$, and $k,l, m \in \mathbb{N}_0$. Then we have
\ba\label{EqGeilePullBackCommuteFormel}
\mathrm{d}^{X^*\nabla}\mleft( X^!\omega \mright)
&=
X^! \mleft( \mathrm{d}^\nabla \omega \mright), \\
\mathrm{d}^{\nabla+D} \omega
&=
\mathrm{d}^\nabla \omega + D \wedge \omega, \label{eqDifferentialSplit}, \\
\mathrm{d}^\nabla \mleft( T \wedge \omega \mright)
&=
\mathrm{d}^\nabla T \wedge \omega
	+ (-1)^m ~ T \wedge \mathrm{d}^\nabla \omega \label{TypischerSplitdesDifferentialsaufdasWedgeProdukt}
\ea
for all $\omega \in \Omega^l(N; K)$, $\psi \in \Omega^k(N;K)$, $D \in \Omega^1(N; \mathrm{End}(K))$, and $T \in \Omega^m(N; \mathrm{End}(K))$.
\newline

When $K$ is additionally an LAB, then we also have
\ba
\mleft( \mathrm{ad} \circ \omega \mright) \wedge \psi
&=
\mleft[ \omega \stackrel{\wedge}{,} \psi \mright]_K, \label{wedgeproduktmitadLambdaergibtLieklammer} \\
X^!\mleft( \mleft[ \omega \stackrel{\wedge}{,} \psi \mright]_K \mright)
&=
\mleft[ X^!\omega \stackrel{\wedge}{,} X^!\psi \mright]_{X^*K}, \label{eqPullbackofLiebracketStuff} \\
\mleft[ \omega \stackrel{\wedge}{,} \psi \mright]_K
&=
- (-1)^{lk} ~ \mleft[ \psi \stackrel{\wedge}{,} \omega \mright]_K, \label{VertauschungsregelForKKlammerAufFormen}\\
\mleft[ \omega \stackrel{\wedge}{,} \mleft[ \omega \stackrel{\wedge}{,} \omega \mright]_K \mright]_K
&=
0 \label{JacobiIdentityForFormBracket}, \\
\mathrm{ad}^* \circ X^!\omega
&=
X^!\mleft( \mathrm{ad} \circ \omega \mright) \label{EqCommutationRelation}
\ea
for all $\omega \in \Omega^l(N; K)$, $\psi \in \Omega^k(N;K)$, and smooth maps $X: M \to N$, where we write $\mathrm{ad}^*$ for the adjoint representation with respect to $\mleft[ \cdot, \cdot \mright]_{X^*K}$.
\end{propositions}

\begin{remarkohne}
\leavevmode\newline
Eq.~\eqref{VertauschungsregelForKKlammerAufFormen} and Eq.~\eqref{JacobiIdentityForFormBracket} are generalizations of similar expressions just using the Lie algebra bracket $\mleft[ \cdot, \cdot\mright]_{\mathfrak{g}}$ of a Lie algebra $\mathfrak{g}$, which basically is the formulation on trivial LABs, see \cite[\S 5, first and second statement of Exercise 5.15.14; page 316]{hamilton}. Eq.~\eqref{TypischerSplitdesDifferentialsaufdasWedgeProdukt} is of course the typical Leibniz rule of the exterior covariant derivative just extended to the wedge-product with $\mathrm{End}(K)$-valued forms, and Eq.~\eqref{EqGeilePullBackCommuteFormel} is a generalization of the well-known $X^! \circ \mathrm{d} = \mathrm{d} \circ X^!$, where $\mathrm{d}$ is the de-Rham differential (we omit to clarify on which manifold; this should be given by the context).
\end{remarkohne}

\begin{proof}
\leavevmode\newline
\indent $\bullet$ Recall that we have the following property of the pullback connection
\bas
\mleft( X^*\nabla \mright)_Y \mleft( X^* \mu \mright)
&=
X^*\mleft( \nabla_{\mathrm{D}X(Y)} \mu \mright)
\eas
for all $Y \in \mathfrak{X}(M)$, smooth maps $X: M \to N$, connections $\nabla$, and $\mu \in \Gamma(K)$, shortly writing as\footnote{Recall that the pull-back of forms is denoted with an exclamation mark.}
\ba\label{eqShortNotationForPullbackConnections}
\mleft( X^*\nabla \mright) \mleft( X^* \mu \mright)
&=
X^*\mleft( \nabla_{\mathrm{D}X} \mu \mright)
=
X^!(\nabla \mu),
\ea
viewing terms like $\nabla \mu$ as an element of $\Omega^1(N; K)$, $\mathfrak{X}(N) \ni \xi \mapsto \nabla_\xi \mu$, such that we can apply Eq.~\eqref{EqPullBackFormelFuerVerschiedeneDefinitionen}. That extends to exterior covariant derivatives by fixing a local frame $\mleft(e_a\mright)_a$ of $K$ (also used in the following), then we have $\omega^a \in \Omega^l(U)$ ($l \in \mathbb{N}_0$) such that locally
\bas
\omega
&=
\omega^a \otimes e_a
\eas
for all $\omega \in \Omega^l(N; K)$. The pull-back of forms clearly splits over this tensor product by its definition, \textit{i.e.}
\bas
X^!\omega
&=
X^!\omega^a \otimes X^* e_a,
\eas
and the exterior covariant derivative is then calculated by
\bas
\mathrm{d}^{X^*\nabla}\mleft( X^!\omega \mright)
&=
\underbrace{\mathrm{d}\mleft( X^! w^a \mright)}_{\mathclap{= X^! \mleft(\mathrm{d} \omega^a\mright)}} \otimes~ X^*e_a
	+ (-1)^l ~ X^!w^a \wedge \underbrace{\mleft( X^*\nabla \mright)\mleft( X^*e_a \mright)}_{\stackrel{\text{Eq.~\eqref{eqShortNotationForPullbackConnections}}}{=} X^!\mleft( \nabla e_a \mright)} \nonumber \\
&=
X^!\mleft( \mathrm{d}\omega^a \otimes e_a + (-1)^l ~ \omega^a \wedge \nabla e_a \mright) \nonumber \\
&=
X^! \mleft( \mathrm{d}^\nabla \omega \mright). 
\eas

$\bullet$ Observe
\bas
\mathrm{d}^{\nabla+D} \omega
&=
\mathrm{d}\omega^a \otimes e_a + (-1)^l ~ \omega^a \wedge \mleft(\nabla + D\mright) e_a
= \mathrm{d}^\nabla \omega + D \wedge \omega
\eas
for all $\omega\in\Omega^l(N;K)$, $D \in \Omega^1(N;K)$, and connections $\nabla$ on $K$.

$\bullet$ Now let $T \in \Omega^m(N; \mathrm{End}(K))$ and $\mleft( L_a \mright)_a$ a frame of $\mathrm{End}(K)$, such that we can write $T= T^a \otimes L_a$, then
\bas
\mathrm{d}^\nabla(T \wedge \omega)
&=
\mathrm{d}^\nabla(T(e_a) \wedge \omega^a)
=
\mathrm{d}^\nabla(T(e_a)) \wedge \omega^a
	+ (-1)^m ~ T(e_a) \wedge \mathrm{d} \omega^a
\eas
for all $\omega \in \Omega^l(N; K)$, and 
\bas
&&
\mleft( \mathrm{d}^\nabla T \mright)(e_a)
&=
\mathrm{d}T^b \otimes L_b(e_a)
	+ (-1)^m ~ T^b \wedge \underbrace{(\nabla L_b)(e_a)}
	_{\mathclap{= ~ \nabla (L_b(e_a)) - L_b(\nabla e_a)}} \\
&&&=
\mathrm{d}^\nabla(T(e_a))
	- (-1)^m ~ T^b \wedge L_b(\nabla e_a) \\
&&&=
\mathrm{d}^\nabla(T(e_a))
	- (-1)^m ~ \underbrace{\mleft(T^b \otimes L_b\mleft( e_c \mright) \mright)}_{= ~ T(e_c)} \wedge ~ \mleft(\nabla e_a\mright)^c \\
&&&=
\mathrm{d}^\nabla(T(e_a))
	- (-1)^m ~ T \wedge \nabla e_a \\
&\Leftrightarrow&
\mathrm{d}^\nabla(T(e_a))
&=
\mleft( \mathrm{d}^\nabla T \mright)(e_a) + (-1)^m ~ T \wedge \nabla e_a.
\eas
Combining both equations, we arrive at
\bas
\mathrm{d}^\nabla(T \wedge \omega)
&=
\mathrm{d}^\nabla T \wedge \omega
	+ (-1)^m ~ T(e_a) \wedge \mleft( \mathrm{d}\omega^a + (-1)^l ~ w^b \wedge \mleft(\nabla e_b\mright)^a \mright) \\
&=
\mathrm{d}^\nabla T \wedge \omega
	+ (-1)^m ~ T \wedge \mathrm{d}^\nabla \omega.
\eas

In the following let $K$ also be an LAB.

$\bullet$ We also have
\bas
&(\underbrace{\mleft( \mathrm{ad} \circ \omega \mright)}_{\mathclap{\in ~ \Omega^l(N;~ \mathrm{End}(K))}} \wedge ~\psi)(Y_1, \dotsc, Y_{l+k}) \\
&\hspace{1cm}\stackrel{\mathclap{\text{Def.~\eqref{DefVonWedgedemitEnd}}}}{=}~~~
\frac{1}{k!l!}
\sum_{\sigma \in S_{k+l}}
\mathrm{sgn}(\sigma) ~
	\mleft[ \omega\mleft(Y_{\sigma(1)}, \dotsc, Y_{\sigma(l)}\mright), \psi\mleft(Y_{\sigma(l+1)}, \dotsc, Y_{\sigma(l+k)} \mright) \mright]_K \\
&\hspace{1cm}\stackrel{\mathclap{\text{Def.~\ref{def:GradingOfProducts}}}}{=}~~~
\mleft[ \omega \stackrel{\wedge}{,} \psi \mright]_K(Y_1, \dotsc, Y_{l+k})
\eas
for all $w\in\Omega^l(N;K)$, $\psi \in \Omega^k(N;K)$, and $Y_1, \dotsc, Y_{l+k} \in \mathfrak{X}(N)$, where $S_{k+l}$ is the group of permutations $\{1, \dotsc, k+l\}$.

$\bullet$ By definition of $X^*K$ we have
\bas
\mleft[ X^*\mu, X^*\nu \mright]_{X^*K}
&=
X^*\mleft( \mleft[ \mu, \nu \mright]_{K} \mright)
\eas
for all smooth maps $X: M \to N$ and $\mu, \nu \in \Gamma(K)$. Let $\mleft( e_a \mright)_a$ be again a fixed frame of $K$, $\omega =  \omega^a \otimes e_a \in \Omega^l(N;K)$ and $\psi = \psi^a \otimes e_a \in \Omega^k(N;K)$, then, again using Def.~\ref{def:GradingOfProducts},
\bas
X^!\mleft( \mleft[ \omega \stackrel{\wedge}{,} \psi \mright]_K \mright)
&=
X^!\mleft( \mleft[ e_a , e_b \mright]_K \otimes \omega^a \wedge \psi^b \mright)
=
\underbrace{X^*\mleft(\mleft[ e_a , e_b \mright]_K\mright)}_{\mathclap{=~\mleft[ X^*e_a, X^*e_b \mright]_{X^*K}}} \otimes X^!\omega^a \wedge X^!\psi^b
=
\mleft[ X^!\omega \stackrel{\wedge}{,} X^!\psi \mright]_{X^*K}.
\eas

$\bullet$ The antisymmetry of the Lie bracket generalizes to
\bas
\mleft[ \omega \stackrel{\wedge}{,} \psi \mright]_K
&=
\underbrace{\mleft[ e_a, e_b \mright]_K}
_{=~ -\mleft[ e_b, e_a \mright]_K} 
\otimes \underbrace{\omega^a \wedge \psi^b}
_{=~(-1)^{lk} \psi^b \wedge \omega^a}
=
- (-1)^{lk} ~ \mleft[ \psi \stackrel{\wedge}{,} \omega \mright]_K
\eas
for all $\omega \in \Omega^l(N;K)$ and $\psi \in \Omega^k(N;K)$.

$\bullet$ Let $\mleft( e_a \mright)_a$ be still a local frame of $K$, then
\bas
&&
\mleft[ \omega \stackrel{\wedge}{,} \mleft[ \omega \stackrel{\wedge}{,} \omega \mright]_K \mright]_K
~~~&\stackrel{\mathclap{\text{Eq.~\eqref{VertauschungsregelForKKlammerAufFormen}}}}{=}~~~
- (-1)^{2l^2} ~
\mleft[ \mleft[ \omega \stackrel{\wedge}{,} \omega \mright]_K \stackrel{\wedge}{,} \omega \mright]_K \\
&&
&=
- \underbrace{\mleft[ \mleft[ e_a, e_b \mright]_K, e_c \mright]_K}
_{\stackrel{\text{Jacobi}}{=}~ \mleft[ e_a, \mleft[ e_b, e_c \mright]_K \mright]_K + \mleft[ e_b, \mleft[ e_c, e_a \mright]_K \mright]_K}
 \otimes ~\omega^a \wedge \omega^b \wedge \omega^c \\
&&
&=
- \mleft[ \omega \stackrel{\wedge}{,} \mleft[ \omega \stackrel{\wedge}{,} \omega \mright]_K \mright]_K
	- \mleft[ e_b, \mleft[ e_c, e_a \mright]_K \mright]_K \otimes 
	\underbrace{\omega^a \wedge \omega^b \wedge \omega^c}_{\mathclap{=~ (-1)^{2l^2} \omega^b \wedge \omega^c \wedge \omega^a}} \\
&&
&=
-2~ \mleft[ \omega \stackrel{\wedge}{,} \mleft[ \omega \stackrel{\wedge}{,} \omega \mright]_K \mright]_K \\
&\Leftrightarrow&
\mleft[ \omega \stackrel{\wedge}{,} \mleft[ \omega \stackrel{\wedge}{,} \omega \mright]_K \mright]_K
&= 0
\eas
for all $\omega \in \Omega^l(N;K)$.

$\bullet$ We also have
\bas
\mleft[ X^!\omega, X^*\mu \mright]_{X^*K}
&\stackrel{\text{Eq.~\eqref{eqPullbackofLiebracketStuff}}}{=}
X^!\mleft( \mleft[ \omega, \mu \mright]_K \mright)
=
X^!\Big( (\mathrm{ad} \circ \omega)(\mu) \Big)
=
\underbrace{\mleft(X^!\mleft( \mathrm{ad} \circ \omega \mright)\mright)}_{\mathclap{\in ~ \Omega^1(M; ~\mathrm{End}(X^*K))}}(X^*\mu)
\eas
for all $\mu \in \Gamma(K)$, $\omega \in \Omega^l(N;K)$, and smooth maps $X: M \to N$, where we used $(X^*T)(X^*\mu) = X^*(T(\mu))$ for all $T \in \Gamma(\mathrm{End}(K))$ for the last equality. Since sections of $X^*K$ are generated by pullbacks of sections of $K$, we can conclude
\bas
\mathrm{ad}^* \circ X^!\omega
&=
X^!\mleft( \mathrm{ad} \circ \omega \mright).
\eas
\end{proof}

When we add the compatibility conditions~\eqref{CondSGleichNullLAB}, then we have a few more identities.

\begin{corollaries}{Identities related to Lie bracket derivations}{IdentitiesFuerBianchiZeugs}
Let $K \to N$ be an LAB, equipped with a connection $\nabla$ satisfying compatibility condition~\eqref{CondSGleichNullLAB}; also let $M$ be another smooth manifold and $X: M \to N$ a smooth map. Then
\ba\label{eqDerivationOfDifferentialOnBracketonK}
\mathrm{d}^\nabla\mleft( \mleft[ \omega\stackrel{\wedge}{,} \psi \mright]_K \mright)
&=
\mleft[ \mathrm{d}^\nabla \omega \stackrel{\wedge}{,} \psi \mright]_K
	+ (-1)^l~ \mleft[ \omega \stackrel{\wedge}{,} \mathrm{d}^\nabla \psi \mright]_K, \\
\mathrm{d}^\nabla \mleft( \mathrm{ad} \circ \omega \mright)
&=
\mathrm{ad} \circ \mathrm{d}^\nabla \omega \label{DifferentialvonNabalVertauschmitAd}
\ea
for all $\omega \in \Omega^l(N; K)$ and $\psi \in \Omega^k(N; K)$.
%
\end{corollaries}

\begin{remarkohne}
\leavevmode\newline
Eq.~\eqref{eqDerivationOfDifferentialOnBracketonK} is a direct generalization of \cite[\S 5, third statement of Exercise 5.15.14 where it is stated for $\mathfrak{g}$ (trivial LAB with canonical flat connection); page 316]{hamilton}.
\end{remarkohne}

\begin{proof}
\leavevmode\newline
\indent $\bullet$ Using compatibility condition~\eqref{CondSGleichNullLAB} and a local frame $\mleft( e_a \mright)_a$ of $K$,
\bas
\mathrm{d}^\nabla \mleft( \mleft[ \omega \stackrel{\wedge}{,} \psi \mright]_K \mright)
&=
\mathrm{d}^\nabla\mleft( \mleft[ e_a, e_b \mright]_K \otimes \omega^a \wedge \psi^b \mright) \\
&=
\underbrace{\nabla \mleft( \mleft[ e_a, e_b \mright]_K \mright)}
_{=~ \mleft[ \nabla e_a, e_b \mright]_K + \mleft[ e_a, \nabla e_b \mright]_K}
	\wedge ~ \omega^a \wedge \psi^b
	+ \mleft[ e_a, e_b \mright]_K \otimes \mathrm{d}\omega^a \wedge \psi^b \\
&\hspace{1cm}
	+ (-1)^l ~ \mleft[ e_a, e_b \mright]_K \otimes \omega^a \wedge \mathrm{d} \psi^b \\
&=
 \mleft[ e_a, e_b \mright]_K \otimes \mleft( \nabla e_c \mright)^a \wedge \omega^c \wedge \psi^b
	+ (-1)^l ~ \mleft[ e_a, e_b \mright]_K \otimes \omega^a \wedge \mleft( \nabla e_c \mright)^b \wedge \psi^c \\
&\hspace{1cm}
	+ \mleft[ e_a, e_b \mright]_K \otimes \mathrm{d} \omega^a \wedge \psi^b
	+ (-1)^l ~ \mleft[ e_a, e_b \mright]_K \otimes \omega^a \wedge \mathrm{d} \psi^b \\
&=
\mleft[ e_a, e_b \mright]_K \otimes \Big(
	\Big( \underbrace{\mleft( \nabla e_c \mright)^a \wedge \omega^c + \mathrm{d} \omega^a}
	_{=~ \mleft( \mathrm{d}^\nabla \omega \mright)^a} \Big) \wedge \psi^b
	+ (-1)^l ~ \omega^a \wedge \mleft( \mleft( \nabla e_c \mright)^b \wedge \psi^c + \mathrm{d} \psi^b \mright)
\Big) \\
&=
\mleft[ \mathrm{d}^\nabla \omega \stackrel{\wedge}{,} \psi \mright]_K
	+ (-1)^l~ \mleft[ \omega \stackrel{\wedge}{,} \mathrm{d}^\nabla \psi \mright]_K
\eas
for all $\omega \in \Omega^l(N;K)$ and $\psi \in \Omega^k(N;K)$.

$\bullet$ Then by Eq.~\eqref{TypischerSplitdesDifferentialsaufdasWedgeProdukt} and~\eqref{wedgeproduktmitadLambdaergibtLieklammer}, we get
\bas
\mathrm{d}^\nabla \mleft( \mleft[ \omega \stackrel{\wedge}{,} \psi \mright]_K \mright)
&=
\mathrm{d}^\nabla \mleft( (\mathrm{ad} \circ \omega) \wedge \psi \mright)
=
\mathrm{d}^\nabla \mleft( \mathrm{ad} \circ \omega \mright) \wedge \psi
	+ (-1)^l ~ (\mathrm{ad} \circ \omega) \wedge \mathrm{d}^\nabla \psi,
\eas
and we can rewrite Eq.~\eqref{eqDerivationOfDifferentialOnBracketonK}
\bas
\mathrm{d}^\nabla \mleft( \mleft[ \omega \stackrel{\wedge}{,} \psi \mright]_K \mright)
&=
\mleft( \mathrm{ad} \circ \mathrm{d}^\nabla \omega \mright) \wedge \psi
	+ (-1)^l ~ (\mathrm{ad} \circ \omega) \wedge \mathrm{d}^\nabla \psi.
\eas
Combining both, we have
\bas
\mathrm{d}^\nabla \mleft( \mathrm{ad} \circ \omega \mright) \wedge \psi
&=
\mleft( \mathrm{ad} \circ \mathrm{d}^\nabla \omega \mright) \wedge \psi
\eas
for all $\omega \in \Omega^l(N;K)$ and $\psi \in \Omega^k(N;K)$. By (locally) using the 0-forms $\psi = e_a$ for all $a$, this implies Eq.~\eqref{DifferentialvonNabalVertauschmitAd}.
\end{proof}

\renewcommand\refname{List of References}

\bibliography{Literatur}

\begin{thebibliography}{1}

\bibitem{hamilton}
Mark~JD Hamilton.
\newblock {\em Mathematical {G}auge {T}heory}.
\newblock Springer, 2017.

\bibitem{mackenzieGeneralTheory}
K.~Mackenzie.
\newblock General {T}heory of {L}ie {G}roupoids and {A}lgebroids.
\newblock {\em London Mathematical Society Lecture Note Series}, 213, 2005.

\bibitem{DaSilva}
Ana~Cannas Da~Silva and Alan Weinstein.
\newblock {\em Geometric models for noncommutative algebras}, volume~10.
\newblock American Mathematical Soc., 1999.

\bibitem{CurvedYMH}
Alexei Kotov and Thomas Strobl.
\newblock Curving {Y}ang-{M}ills-{H}iggs gauge theories.
\newblock {\em Physical Review D}, 92(8):085032, 2015.

\bibitem{basicconn}
Camilo~Arias Abad and Marius Crainic.
\newblock Representations up to homotopy of {L}ie algebroids.
\newblock {\em Journal f{\"u}r die reine und angewandte Mathematik (Crelles
  Journal)}, 2012(663):91--126, 2012.

\end{thebibliography}
\bibliographystyle{unsrt}


\end{document}